\documentclass[12pt]{article}
\usepackage{amsmath}
\allowdisplaybreaks[4]%¹«Ê½¿çÒ³
\usepackage{amssymb}
\usepackage{amsthm}
\usepackage{fullpage}
\usepackage[utf8]{inputenc}
\usepackage{mathtools}
\usepackage{url}
\usepackage[usenames,svgnames]{xcolor}
\usepackage{cite}
\usepackage[title,titletoc,toc]{appendix}
\usepackage[T1]{fontenc}%¼Ó×÷Õß
\usepackage[utf8]{inputenc}%¼Ó×÷Õß
\usepackage{authblk}%¼Ó×÷Õß

\usepackage[pagebackref=false]{hyperref} %hyperref wants to be loaded last. Set = false if no pagebackref needed
\usepackage[all]{hypcap} %fix hyperref links to tables/figures/etc.
\theoremstyle{plain}
\newtheorem{THM}{Theorem}[section]

\newtheorem{LEM}{Lemma}[section]
\newtheorem{COR}{Corollary}[section]

\theoremstyle{definition}

\theoremstyle{definition}

\numberwithin{equation}{section} % requires package amsthm

\def\lvs{\left <0\right |}
\def\rvs{\left |0\right >}
\def\lvsn{\left <n\right |}
\def\rvsn{\left |n\right >}
\def\lbr{\left <}
\def\rbr{\right >}
\def\normord{\scriptstyle{\bullet\atop\bullet}}
\def\bt{\mathbf{t}}
\def\bs{\mathbf{s}}

\title{2D Toda $\tau$ Functions, Weighted Hurwitz Numbers and the Cayley Graph:  Determinant Representation and Recursion Formula}
\author{Xiang-Mao DING}
\author{Xiang Li \thanks{Lxiang $@$ amss.ac.cn}}
\affil{Institute of Applied Mathematics\authorcr
Academy of Mathematics and Systems Science\authorcr
Chinese Academy of Sciences}

\date{} % È¥µôÈÕÆÚ

\begin{document}
  \maketitle
\baselineskip 16pt

%%%%%%%%%%%%%%%% ?a¨°a %%%%%%%%%%%%%%%%
\begin{abstract}\par
We generalize the determinant representation of the KP $\tau$ functions to the case of the 2D Toda $\tau$ functions. The generating functions for the weighted Hurwitz numbers are a parametric family of 2D Toda $\tau$ functions; for which we give a determinant representation of weighted Hurwitz numbers. Then we can get a finite-dimensional equation system for the weighted Hurwitz numbers $H^d_{G}(\sigma,\omega)$ with the same dimension $|\sigma|=|\omega|=n$. Using this equation system, we calculated the value of the weighted Hurwitz numbers with dimension $0,\,1,\,2$ and give a recursion formula to calculating the higher dimensional weighted Hurwitz numbers. For any given weighted generating function $G(z)$, the weighted Hurwitz number degenerates into the Hurwitz numbers when $d=0$. We get a matrix representation for the Hurwitz numbers. The generating functions of weighted paths in the Cayley graph of the symmetric group are a parametric family of 2D Toda $\tau$ functions; for which we obtain a determinant representation of weighted paths in the Cayley graph.
\end{abstract}
%%%%%%%%%%%%%%%%% Contents %%%%%%%%%%%%%%%%%
%\tableofcontents

%%%%%%%%%%%%%%%% Section 1. %%%%%%%%%%%%%%%%
\section{Introduction: 2D $\tau$-functions and weighted Hurwitz numbers}
\label{chap:introduction}
~\par
Nonlinear integrable systems are one of the most important topics in modern mathematical physics. Two special ones are: Kadomtsev-Petviashvili (KP) and 2D systems. Their $\tau$ functions are especially interesting, due to their being involved in various applications including topology and group representation theory. The KP $\tau$ functions and 2D Toda $\tau$ functions have been studied extensively. In the 1980s, the Kyoto school used free fermions to construct the KP and 2D Toda $\tau$ functions \cite{DJKM,MJ_TMSoliton,MJD,KM}. In this way, the $\tau$ functions could be expressed as the product of the vacuum expectation value of the group-like elements. The group-like elements belong to an infinite dimensional Clifford algebra generated by the free fermions. The KP hierarchies are an infinite set of partial differential equations, with an infinite number of independent variables $\{\bt\}=\{t_1,\,t_2,\, t_3,\, \cdots\}$. 2D Toda hierarchies are an infinite set of partial differential equations, with an infinite number of independent variables $\{\bt\}=\{t_1,\,t_2,\, t_3,\, \cdots\}$ and $\{\bs\}= \{s_1,\,s_2,\, s_3,\, \cdots\}$\cite{UT,Takebe}. In the $\bs=0$ case, the 2D Toda $\tau$ functions degenerate to the KP $\tau$ functions. In the 1990s, using the boson-fermion correspondence, and using the generalized Wick theorem,  S. Kharchev obtained the determinant representation of the KP $\tau$ function \cite{Ka}. \par

Hurwitz numbers $H(\mu^{(1)}, \dots , \mu^{(k)})$ are the inequivalent $ N$-sheeted branched covers of the Riemann sphere\cite{HurwitzUber,HurwitzUber2}, with specified classes of ramification profile structures $\{\mu^{(1)}, \dots,$ $ \mu^{(k)}\}$. In combinatorial terms, the Hurwitz numbers have an equivalent definition\cite{Frobenius1,Frobenius2}: the number of distinct factorizations of the identity element $I_N=h_1h_2\cdots h_k$ in the symmetry group $\mathbf{S}_N$ divided by the normalization factor $N!$. The factorization elements $h_i$ have conjugate class $cyc(\mu^{i})$. The weighted Hurwitz numbers $H^d_{G}(\mu,\nu)$ are the inequivalent weighted branched covers of the Riemann sphere\cite{HR}. The weighted branched covers have branch points at $0$ and $\infty$ with ramification profiles $\mu,\,\nu$, and have distinct, finite, non-zero branch points with ramification profiles $\{\mu^1,\cdots,\mu^k\}$. The weight of the weighted branched covers are related to a weight generating function $G(z)$ of $H^d_{G}(\mu,\nu)$. And the sum of the co-lengths of those ramification profiles $\mu^1,\cdots,\mu^k$ is $d$ in $H^d_{G}(\mu,\nu)$. In the $d=0$ case, the weighted Hurwitz numbers $H^d_{G}(\mu,\nu)$ degenerate to the Hurwitz numbers $H(\mu,\nu)$. After choosing an appropriate weighted generating function, the weighted Hurwitz numbers\cite{PJGen,PJ2D} can cover the various types of Hurwitz numbers discussed in\cite{AMMN,GGN,GGN2,KZ,OToda,P}. M .Guay-Paquet and J. Harnad investigated a parametric family of 2D Toda $\tau$ functions \cite{PJGen,PJ2D}, which are the generating functions of the weighted Hurwitz numbers and weighted paths in the Cayley graph of the symmetric group $\mathbf{S}_n$. So we are able to study the weighted Hurwitz numbers using the 2D Toda $\tau$ functions. A. Alexandrov, G. Chapuy, B. Eynard and J. Harnad found a topological recursion of weighted Hurwitz numbers by applying recursion to the form $\tilde{\omega}_{g,n}$ defined in terms of the correlators $\tilde{\mathcal{W}}_{g,n}$ \cite{ACEH,ACEH2}. The correlators $\tilde{\mathcal{W}}_{g,n}$ are the derivative of $\tilde{F}_{g,n}(\bs,x_1,x_2,\cdots,x_n)$ which is another type of generating function for the weighted Hurwitz numbers. The 2D Toda $\tau$ functions have two sets of independent variables $\{\bt,\bs\}$,they allow the 2D Toda flows $e^{J_- ({\mathbf s})}$ to be treated as other group-like elements, when deriving the properties of the weighted Hurwitz numbers using the 2D Toda $\tau$ functions. Therefore, all the operators depends on the variables $\{\bs\}$.\par
In this paper, we generalize the determinant representation of the KP $\tau$ functions to the case of the 2D Toda $\tau$ functions
\begin{equation}\nonumber
\begin{split}
\tau^{2D}(\bt,\mathbf{s})=\frac{\lbr-N\right| G\left|-N\rbr}{\Delta(\nu)\Delta(\mu)\nu_1\cdots\nu_N\mu_1\cdots\mu_N}\mathop{det}\limits_{i,j=1\cdots N}\Big(\frac{\lbr-N\right|\psi^*(\nu_j)G\psi(\mu_i^{-1})\left|-N\rbr}{\lbr-N\right| G\left|-N\rbr}\Big).
\end{split}
\end{equation}
Here we take the naive approach of treating 2D Toda $\tau$ functions as a function of $\bt$ and $\bs$, and we treat the boson-fermion correspondence to the dual 2D Toda flows $e^{J_+ ({\mathbf t})}$ and $e^{J_- ({\mathbf s})}$ in the same way. At the same time, we leave $G$ as a group-like element. Using the corollary of the generalized Wick theorem, we get the determinant representation of the 2D Toda $\tau$ functions.\par
The generating functions for the weighted Hurwitz numbers
are a parametric family of 2D Toda $\tau$ functions. With this result,  we get a determinant representation of the weighted Hurwitz numbers,
\begin{equation}\label{Int0}
\sum_{d=0}^\infty \beta^d \sum_{\substack{\omega,\sigma\\|\omega|=|\sigma|}}H^d_G(\omega,\sigma)p_\omega(\bt)p_\sigma(\mathbf{s})
=\frac{r_0(-N)}{\Delta(\nu)\Delta(\mu)}\mathop{det}\limits_{i,j=1\cdots N}\Big(\sum_{l=-N}^\infty \nu_j^{-l-1}\mu_i^{-l-1}\rho_l\Big),
\end{equation}
where the $\bt$, $\mathbf{s}$ are Miwa parameters $t_k=\frac{1}{k}\sum_{i=1}^{N}\nu_i^{-k},\, s_l=\frac{1}{l}\sum_{j=1}^{M}\mu_j^{-l}$ \cite{Miwa}. If we consider the terms of ${\nu ^{-n}  \mu ^{-n}}$, which means the sum of the power of $\{\mu_i\}_{i=1}^N$ or $\{\nu_j\}_{j=1}^N$ are $-n$, we could get a finite-dimensional equation system for the weighted Hurwitz numbers of the same dimension $|\omega|=|\sigma|=n$. In the $N<n$ case, it is an underdetermined system of equations. And for $\forall N,\,N\geq n$, we could get the same appropriate system of equations. That is to say, we get the hierarchies of weighted Hurwitz numbers for dimension $n$.
We calculate the value of the weighted Hurwitz numbers when $n$ is $0,\,1,\,2,\,3$. But as the dimension of the weighted Hurwitz numbers increases, the calculation becomes more and more complicated. We give a recursion formula to calculating the higher dimensional weighted Hurwitz number hierarchies.
\begin{equation}\nonumber
\begin{split}
   \sum_{d=0}^\infty &  \beta^d \sum_{\substack{\omega,\sigma\\|\omega|=|\sigma|=k+1}}H^d_G(\omega,\sigma) p_\omega(\bt)p_\sigma(\mathbf{s})\\
= & \frac{\rho_{k}}{\rho_{-1}}\frac{\sum\limits_{\zeta\in S_{k+1}}sgn(\zeta)\mu_{\zeta(1)}^k \cdots\mu_{\zeta(k)}\mu_{\zeta(k+1)}^{-k-1}} {\Delta(\mu)}\frac{\sum\limits_{\tilde{\zeta}\in S_{k+1}}sgn(\tilde{\zeta})\nu_{\tilde{\zeta}(1)}^k \cdots\nu_{\tilde{\zeta}(k)}\nu_{\tilde{\zeta}(k+1)}^{-k-1}}{\Delta(\nu)} \\
  & + \sum_{l_{2}=-1}^{k-1}\frac{\rho_{l_{2}}}{\rho_{-2}} \sum_{\substack{\omega,\sigma\\|\omega|=|\sigma|=k-l_{2}-1}} \bar{p}_{\sigma}(\bt,k-1,l_2) \bar{p}_{\omega}(\bs,k-1,l_2) \sum_{d=0}^\infty {\beta}^d Term_{(\rho_j,\,j\geq l_{2}+1)}\Big\{H^d_G(\omega,\sigma)\Big\}\\
  & + \cdots\\
  & + \sum_{l_{k+1}=-k}^{0}\frac{\rho_{l_{k+1}}}{\rho_{-k-1}} \sum_{\substack{\omega,\sigma\\|\omega|=|\sigma|=-l_{k+1}}} \bar{p}_{\sigma}(\bt,1,l_{k+1})\bar{p}_{\omega}(\bs,1,l_{k+1}) \sum_{d=0}^\infty {\beta}^d r_0(-k) Term_{(\rho_j,\,j\geq l_{k+1}+1)}\Big\{\frac{H^d_G(\omega,\sigma)} {r_0(-k)}\Big\},
\end{split}
\end{equation}
where $Term_{(\rho_j,\,j\geq l_{N+1}+1)}\{f(\rho_j)\}$ denotes the terms $\rho_{l_1},\,\cdots,\,\rho_{l_N}$, whose subscripts $\{l_1,\,\cdots,\,l_{N}\}$ are greater than or equal to $l_{N+1}+1$, in the function $f(\rho_j)$ which is a polynomial with respect to $\{\rho_j\}_{j\in\mathbb{Z}}$. \par
$\bar{p}_{\sigma}(\bt,m,l)$ and $\frac{1}{\Delta(\mu)}\sum\limits_{\zeta\in S_{k+1}}sgn(\zeta)\mu_{\zeta(1)}^k \cdots$ $\mu_{\zeta(k)}\mu_{\zeta(k+1)}^{-k-1}$ could be written as a linear combination of $p_\lambda(\bt)$, while $\{p_\lambda(\bt)\}$ are a set of bases of the symmetric function ring $\Lambda(\bt)$. If we could get the coefficients of the linear combinations of $p_\lambda(\bt)$ and $p_\lambda(\bs)$, we could find the recursion formula for the weighted Hurwitz numbers, which means finding the value of $H^d_G(\omega,\sigma),\,|\omega|=|\sigma|=k+1$ from the value of $H^d_G(\omega,\sigma),\,|\omega|=|\sigma|<k+1$.\par
The determinant representation of the weighted Hurwitz numbers (\ref{Int0}) could also be expanded in terms of $\beta$. For any weighted generating function $G(z)$, the weighted Hurwitz number degenerates into the Hurwitz numbers when $d=0$. The zeroth-order expansions of $G(z),\,\rho_j,\,r_0(-n)$ are 1, and we have
\begin{equation}\nonumber
\begin{split}
  \sum_{\substack{\omega,\sigma\\|\omega|=|\sigma|}}H^0_G(\omega,\sigma) p_\omega(\bt)p_\sigma(\mathbf{s})
= & \prod\limits_{1\leq i,j\leq n}\sum\limits_{k_{ij}=-\infty}^{0}\nu_j^{k}\mu_i^{k}\\
= & \sum_{k_{11}\cdots k_{nn}=-\infty}^{0}\prod\limits_{1\leq j\leq n}\nu_j^{\sum_{i=1}^{n}k_{ij}}\prod\limits_{1\leq i \leq n}\mu_i^{\sum_{j=1}^{n}k_{ij}}.
\end{split}
\end{equation}
and $p_\omega(\bt),\,p_\sigma(\mathbf{s}),\prod\limits_{1\leq j\leq n}\nu_j^{\sum_{i=1}^{n}k_{ij}}\prod\limits_{1\leq i \leq n}\mu_i^{\sum_{j=1}^{n}k_{ij}}$ could be decomposed into the sum of $n\times n$-dimensional matrix groups with non-positive integer coefficients. In this way, we obtain the matrix representation of the Hurwitz numbers. We could work out the value of the Hurwitz numbers by matrix operations.\par
M .Guay-Paquet and J .Harnad investigated a parametric family of 2D Toda $\tau$ functions \cite{PJGen,PJ2D}, in which are the generating functions of the weighted Hurwitz numbers, and the weighted paths are given through the Cayley graph of the symmetric group $\mathbf{S}_n$. We get the determinant representation of the weighted paths in the Cayley graph of the symmetric group $\mathbf{S}_n$.\par
\begin{align}\nonumber
\sum_{\substack{\omega,\sigma\\|\omega|=|\sigma|=n}}  & m^\lambda_{\mu\nu}p_\omega(\bt) p_\sigma(\mathbf{s})\\
= & \frac{n!}{\Delta(\nu)\Delta(\mu)} \sum_{\substack{l_1,\cdots,l_{n}=-n\\l_1>\cdots>l_n\\l_1+\cdots +l_n=\frac{-n(n-1)}{2}}}^{\infty} (l_1,l_1-1, \cdots,0,l_2,l_2-1,\cdots,\notag\\
  & -1,\cdots,l_n,l_n-1,\cdots,-n+1)^{\lambda}\beta^d\notag\\
  & \times \sum_{\sigma\in S_k} sgn(\sigma)\left|\begin{array}{cccc}
\nu_{\sigma(1)}^{-l_1-1}\mu_{1}^{-l_1-1} & \nu_{\sigma(2)}^{-l_2-1}\mu_{1}^{-l_2-1} & \cdots & \nu_{\sigma(N)}^{-l_n-1}\mu_{1}^{-l_n-1}\\
\nu_{\sigma(1)}^{-l_1-1}\mu_{2}^{-l_1-1} & \nu_{\sigma(2)}^{-l_2-1}\mu_{2}^{-l_2-1} & \cdots & \nu_{\sigma(N)}^{-l_n-1}\mu_{2}^{-l_n-1}\\
\vdots & \vdots & \ddots & \vdots\\
\nu_{\sigma(1)}^{-l_1-1}\mu_{N}^{-l_1-1} & \nu_{\sigma(2)}^{-l_2-1}\mu_{N}^{-l_2-1} & \cdots & \nu_{\sigma(N)}^{-l_n-1}\mu_{N}^{-l_n-1}\\
\end{array}\right|,\notag
\end{align}
in which $m^\lambda_{\mu\nu}$ is the number of monotonic $\lambda$ signature paths in the Cayley graph from $cyc(\mu)$ to $cyc(\nu)$.\par
This paper is organized as follows. In Section 2, we list the basic notations of free fermions, 2D Toda $\tau$ functions and weighted Hurwitz numbers. In the next Section, we generalize the determinant representation of the KP $\tau$ functions to the case of the 2D Toda $\tau$ functions. Then, we give a determinant representation of the weighted Hurwitz numbers and discuss the hierarchies of weighted Hurwitz numbers in Section 4. And we calculate the value of the weighted Hurwitz numbers when $n$ is $0,\,1,\,2,\,3$ and give a recursion formula to calculating the higher dimensional weighted Hurwitz numbers in Section 5. In the following two Sections, we first expand the determinant representation of the weighted Hurwitz numbers along $\beta$, such that we get the matrix representation of the Hurwitz numbers, and then in Section 7, we obtain the determinant representation of the weighted paths in the Cayley graph.
%%%%%%%%%%%%%%%% Subsection 1.1 %%%%%%%%%%%%%%%%
\section{Free fermions, 2D Toda $\tau$ functions and weighted Hurwitz numbers}
\label{chap:Preliminaries}
~\par
In this section, we briefly introduce the free fermions, the 2D Toda $\tau$ functions and the weighted Hurwitz numbers.
\subsection{Free fermions and the generalized Wick  theorem}
~\par
%The notation and proof for details in this section could be found in\cite{AAFree,HB}.\par
Free fermions $\psi_n,\,\psi^*_n,\,n\in\mathbb{Z}$ are anti-commutation operators:
\begin{equation}\label{eq:exchang fermions}
[\psi_n , \psi_m ]_+ = [\psi^*_n, \psi^*_m]_+=0, \quad [\psi_n , \psi^*_m]_+=\delta_{n,m}.
\end{equation}
Free fermions generate an infinite-dimensional Clifford algebra $\mathcal{A}$:
\begin{equation}\nonumber
\begin{split}
\mathcal{A}= & span \{\psi_{m_1}\cdots\psi_{m_r}\psi^*_{n_s}\cdots\psi^*_{n_1}\bigg{|}0>m_1>\cdots>m_r,0\leq n_s<\cdots<n_1,\quad r,s\in\mathbb{Z}\quad r,s\geq0 \}
\end{split}
\end{equation}
and fermion fields are
\begin{equation}\label{fermion fleid1}
\psi (z)=\sum\limits_{k\in \mathbb{Z}}\psi_k z^{k} ,\quad \quad
\psi^* (z)=\sum\limits_{k\in \mathbb{Z}}\psi^*_k z^{-k-1}.
\end{equation}\par
Note that in \cite{AAEnumerative,AAFree,HB}, the fermion fields are defined as $\varPsi^* (z)=\sum\limits_{k\in \mathbb{Z}}\varPsi^*_k z^{-k}$.
The zero vectors $\lvs,\rvs$ are also called the "Dirac sea", which means the states are empty for all states of $n<0$ and occupied for the states $n>0$. Zero vectors have the annihilation relationships:
\begin{equation}\label{annihilation}
\psi_n\rvs= \lvs \psi^*_n =0, \quad n< 0; \qquad   \lvs\psi_m=\psi^*_m\rvs=0,\quad m\geq 0.
\end{equation}\par
The fermionic Fock space $\mathcal{F}$ is a vector space generated by the following vectors:
\begin{equation}\nonumber
\begin{split}
\mathcal{F}= & \{ \psi_{m_1}\cdots\psi_{m_r}\psi^*_{n_s}\cdots\psi^*_{n_1}\rvs\bigg{|}0>m_1>\cdots>m_r, 0\leq n_s<\cdots<n_1,\quad r,s\in\mathbb{Z}\quad r,s\geq0\}.
\end{split}
\end{equation}
For a partition $\lambda=(\lambda_1,\lambda_2,\cdots,\lambda_{l}),\, \lambda_1>\lambda_2>\cdots>\lambda_{l}>0$, where the sum of the non-zero
 $ \lambda_i$ parts is the weight of $\lambda$:
 $|\lambda|=\sum\limits_{i=1}^l \lambda_i$ and length $l(\lambda)=l$, we could write a set of canonical orthonormal bases for the fermionic Fock space:
\begin{equation}\label{lambda_q}
\begin{array}{l}
\left|\lambda,q\rbr :=\psi^*_{q-\beta_1 -1}\cdots \psi^*_{q-\beta_{d(\lambda )}\! -1}\, \psi_{q+\alpha_{d(\lambda )}}\cdots \psi_{q+\alpha_1}\left|q\rbr
\\   \\
\lbr \lambda,q\right|:=
\lbr q\right| \psi^*_{q+\alpha_1}\cdots \psi^*_{q+\alpha_{d(\lambda )}}\, \psi_{q-\beta_{d(\lambda )}\! -1}\cdots \psi_{q-\beta_1 -1},
\end{array}
\end{equation}
where $(\vec \alpha |\vec \beta )=(\alpha_1, \ldots , \alpha_{d(\lambda )}|\beta_1 , \ldots , \beta_{d(\lambda )})$ is the Frobenius notation for a partition $\lambda$ with $\alpha_i =\lambda_i -i,\,\beta_i =\lambda'_i -i$, $\lambda'$ is the transposed partition of $\lambda$, $d(\lambda )$ is the number of  boxes in the main diagonal of $\lambda$, and $q$ is the charge of $\left |\lambda,q\rbr$ or $\lbr \lambda,q\right|$. We don't repeat the details of the proofs, which can be found in \cite{AAFree}. The partition function $p(n)$ is the number of partitions of weight $n$\cite{Andrews}.
The vacuum expectation $\lvs\cdots\rvs$ is the Hamiltonian linear form of the Clifford algebra, which satisfies
\begin{align}
\left<0|0\right > & =1,\\
\lvs(g_1+g_2)\rvs & =\lvs g_1\rvs+\lvs g_2\rvs,\\
\lvs \kappa g_1\rvs & =\kappa\lvs g_1.\rvs
\end{align}\par
According to the vacuum expectation, we could define the Normal Ordering $\normord \big(\displaystyle{\cdots} \big)\normord $ as follows: move all the operators when acting on $\rvs$ give zero to the right, move all the operators when acting on $\lvs$ give zero to the left, and multiply by the factor (-1) for exchanging a pair of fermions.
Using the normal ordering, we can introduce the boson current:
\begin{equation}
J(z)=\normord \displaystyle{\psi(z)\psi^*(z)}\normord\displaystyle{= \sum\limits_{k\in\mathbb{Z}}J_{k}z^{-k}},     %\displaystyle ·Å´ó
\end{equation}
where $J_k=\sum\limits_{j\in \mathbb{Z}}\normord \big(\displaystyle{\psi_j \psi^* _{j+k}}  \big)\normord$ are the bosons. When $k\neq0$, $J_k=\sum\limits_{j\in \mathbb{Z}}\displaystyle\psi_j \psi^* _{j+k}$. When $k=0$, we call $J_0$ the charge operator $Q$, as $Q\left |\lambda,q\right > =q\left |\lambda ,q\right >$.
The commutation relations for bosons are obvious from the commutation relations for fermions:
\begin{equation}\label{exchang Boson}
[J_k,J_l]=k\delta_{k+l,0},\quad[J_k , \psi _m]=\psi_{m-k} ,\quad[J_k ,\psi^* _m]=-\psi^*_{m+k}.
\end{equation}
The bosons generate the Weyl algebra. We use the notation $J_+(\mathbf{t})=\sum\limits_{k\geq 1}t_k J_k$, $J_-(\mathbf{t})=\sum\limits_{k\geq 1}t_k J_{-k}$.\par
The KP/2D Toda flows are defined as
\begin{equation}\label{def:KP-flow}
\hat{\gamma}_+({\mathbf t})=e^{J_+({\mathbf t})},\qquad\hat{\gamma}_-({\mathbf t})=e^{J_-({\mathbf t})}.
\end{equation}
For the bilinear combination of fermions $\sum_{mn} b_{mn}\psi^*_m \psi_n$, the matrix $b = (b_{mn})$ generates an infinite dimensional Lie algebra \cite{MJ_TMSoliton}. The rotation matrix $R=e^{b}$ corresponds to the group elements. It denotes an infinite-dimensional group $(GL(\infty))$ with elements
\begin{equation}\label{gl}
G=\exp \Bigl (\sum_{i, k \in {\mathbb{z} }}b_{ik}\psi^*_i \psi_k\Bigr ),
\end{equation}
called the group-like elements. For the group-like elements $G$, they have the BBC (basic bilinear condition)\cite{AAFree}
\begin{equation} \label{BBC}
\sum_{k \in {\mathbb Z}} \lbr U \right|  \psi_{k} G \left|V \rbr
 \lbr U^{\prime} \right|  \psi^*_{k} G \left|V^{\prime} \rbr =
\sum_{k \in {\mathbb Z}} \lbr U \right| G \psi_{k} \left|V \rbr
 \lbr U^{\prime} \right| G \psi^*_{k} \left|V^{\prime} \rbr,
\end{equation}
where $\left|V \rbr , \left|V^{\prime} \rbr$ and $\lbr U \right|, \lbr U^{\prime} \right|$ are any elements in the Fock space and dual Fock space. Now, we could define the generalized group-like elements as any element of the Clifford algebra $\mathcal{A}$, which satisfies the BBC, thus fermions are generalized group-like elements. The group-like elements we use below mean generalized group-like elements.
\begin{THM}[Generalized Wick theorem]$v_i =\sum_j v_{ij}\psi_j$ ( $ resp.$ $w_i^*=\sum_j w^{*}_{ij}\psi^*_j$) is any linear combination of $\psi_j$ ( $resp.$ $\psi^*_j$), $G, G'$ are group-like elements with zero charge, and for any $n\in\mathbb{Z}$, $\lvsn G'G\rvsn \neq 0$, then
\begin{equation}\label{Wick1}
\frac{\lvsn G' v_1 \ldots v_m w^{*}_m \ldots w^{*}_1 G\rvsn }{\lvsn G'G\rvsn }=\det_{i,j =1,\ldots , m}\frac{\lvsn G'v_j w^{*}_iG\rvsn }{\lvsn G'G\rvsn }.
\end{equation}
\end{THM}
\begin{COR}
\begin{equation}\label{Wick2b}
\frac{\lvsn G' v_1 \ldots v_m G'' w^{*}_m \ldots w^{*}_1 G\rvsn }{\lvsn G'G''G\rvsn }=\det_{i,j =1,\ldots , m}\frac{\lvsn G'v_j G''w^{*}_iG\rvsn }{\lvsn G'G''G\rvsn },
\end{equation}
where $G''$ is also a group-like element. We skip the proofs which can be found in \cite{AAFree}.\par
\end{COR}
Without loss of any generality, the group-like elements in the following have zero charge.
\subsection{Bosonic Field}
~\par
Before introducing the chiral bosonic field $\phi(z)$, we introduce the shift operator $e^{P}=\sum_{q,\lambda}\left|\lambda,q\rbr\lbr\lambda,q-1\right|$. For the definition of the shift operator, we take
\begin{equation}
e^{\pm P}\left|\lambda,q\rbr =\left|\lambda,q\pm1\rbr,\qquad
\lbr\lambda,q\right|e^{\pm P}=\lbr\lambda,q\mp1\right|.
\end{equation}
And we can get the commutation relation between the charge operator and shift operator $[Q,P]=1$ by direct calculation.\par
Assuming that the operators $\varphi_k$, $k\in\mathbb{Z}/\{0\}$ are the elements in the Clifford algebra $\mathcal{A}$, $\varphi_k$ is a linear combination of $\psi_n,\,\psi^*_{n}$. The operators $\varphi_k$ have the following commutation relations
\begin{equation}\label{exchange:varphi P Q}
[\varphi_k,\varphi_l]=k\delta_{k+l,0},\qquad[\varphi_k,P]=0,\qquad[\varphi_k,Q]=0,
\end{equation}
and the annihilation relations
\begin{equation}\label{annihilate varphi}
\begin{split}
\lvs \varphi_{-k} & =0   ,\quad\qquad ~  \lvs \varphi_k\neq0,  \\
\varphi_k \rvs & =0    ,\quad\qquad   \varphi_{-k} \rvs\neq0,         \qquad\qquad      k>0.
\end{split}
\end{equation}
Recall the definition of the normal ordering in the last section is that move all the operators when acting on $\rvs$ give zero to the right, move all the operators when acting on $\lvs$ give zero to the left, and use the commutative relation of operators to exchange each pair of operators. The normal ordering $\normord\scriptstyle{(\displaystyle{\cdots})}\normord$ of $\varphi_{k},\,Q,\,P$ which means move $\varphi_{-k},\,k>0$ to the left, move $\varphi_{k},\,k>0$ to the right, put $Q,\,P$ in the middle of $J_{-k}$ and $J_{k}$ with $P$ at left of $Q$ and use $[Q,P]=1,\, [\varphi_k,\varphi_l]=k\delta_{k+l,0},\,[\varphi_k,P]=0,\,[\varphi_k,Q]=0$ to exchange each pair of these operators. Neither the charge operator nor the shift operator acting on $\lvs$ or $\rvs$ are zero, so it is not necessary to change the positions of them according to the defining principle of the normal order. As Q and P are not interchangeable, without loss of generality, it is an artificial requirement that $\normord\scriptstyle{(\displaystyle{\cdots})}\normord$ makes P on the left of Q. If we require $\normord\scriptstyle{(\displaystyle{\cdots})}\normord$ to make P on the right of Q, all the conclusions of this article will not be changed. \par
Now we can introduce the chiral bosonic field $\phi(z)$:
\begin{equation}\label{def:bosfld1}
\phi(z)=\sum_{k\neq 0}(\frac{\varphi_{-k}}{k}z^k)+Qln(z)+P,
\end{equation}
with
\begin{equation}\label{fermfld corr bosfld1}
\begin{split}
\normord\displaystyle{e^{\phi(z)}}\normord & =\psi(z)=\sum_{i}\psi_iz^i,    \\
\normord\displaystyle{e^{-\phi(z)}}\normord & =\psi^*(z)=\sum_{j}\psi_j^*z^{-j-1}.
\end{split}
\end{equation}
A solution satisfying the equations (\ref{exchange:varphi P Q}),(\ref{annihilate varphi}),(\ref{fermfld corr bosfld1}) exists. We will find one of them below:
\begin{equation}\label{mid 1}
\begin{split}
&  \normord\displaystyle{\exp^{\alpha\phi(z)}}\normord~\normord\displaystyle {\exp^{\beta\phi(w)}}\normord\\
= &  e^{\alpha\sum\limits_{k=1}^{\infty}\frac{\varphi_{-k}} {k}z^k}e^{\alpha P}z^{\alpha Q}e^{-\alpha\sum\limits_{k=1}^{\infty}\frac{\varphi_{k}}{k}z^{-k}} e^{\beta\sum\limits_{k=1}^{\infty} \frac{\varphi_{-k}}{k}w^k}e^{\beta P}w^{\beta Q}e^{-\beta\sum\limits_{k=1} ^{\infty}\frac{\varphi_{k}}{k}w^{-k}}\\
= &  \normord\displaystyle{\exp^{\alpha\phi(z)+\beta\phi(w)}}\normord e^{-\alpha\beta\sum\limits_{k,l=1}^{\infty}\frac{1}{kl}[\varphi_k,\varphi_l] \frac{w^l}{z^k}}e^{\alpha\beta[Qlnz,P]}\\
= &  (z-w)^{\alpha\beta}~\normord\displaystyle{\exp^{\alpha\phi(z)+\beta\phi(w)}} \normord.
\end{split}
\end{equation}
Next, let $w$ approximate $z$, and always ensure that the norm of $w$ is smaller than $z$'s during the approximation
\begin{equation}\label{mid 2}
\begin{split}
&  \psi(z)\psi^*(z)\\
= &  \lim_{w\rightarrow z}\psi(z)\psi^*(w)\\
= &  \lim_{w\rightarrow z}\frac{\normord \displaystyle{(\exp^{\phi(z)-\phi(w)})}\normord }{z-w}\\
= &  \normord \displaystyle{\lim_{w\rightarrow z}(\frac{1}{z-w}+\frac{\phi(z)-\phi(w)}{z-w}+\frac{1}{2}\frac{(\phi(z)-\phi(w))^2} {z-w}+\cdots)}\normord \\
= &  \normord \displaystyle{\lim_{w\rightarrow z}\frac{1}{z-w}+\partial_z\phi(z)}\normord \\
= &  \lim_{w\rightarrow z}\frac{1}{z-w}+\frac{1}{z}\sum\limits_{k\neq0} (\varphi_{-k}z^{k})+\frac{Q}{z}.
\end{split}
\end{equation}
Notice that $\psi(z)\psi^*(w)-\normord\displaystyle{ \psi(z)\psi^*(w)}\normord=\frac{1}{z-w}$, substituting this into (\ref{mid 2}) yields
\begin{equation}\nonumber
\normord\displaystyle{\psi(z)\psi^*(z)}\normord \displaystyle{=\frac{1}{z}\sum_{k\neq0}(\varphi_{-k}z^k)+\frac{Q}{z}}
\end{equation}
and
\begin{equation}\label{fer-J}
\normord\displaystyle{\psi(z)\psi^*(z)}\normord \displaystyle{=\sum\limits_{i,j\in\mathbb{Z}}}\normord\displaystyle{\psi_i\psi_{i+j}^*}\normord \displaystyle{z^{-j-1}}=\displaystyle{\sum_{j\in\mathbb{Z}}\frac{J_i}{z^{j+1}}}.
\end{equation}
Noting that $J_0=Q$, we find a set of solutions for the system of equations (\ref{exchange:varphi P Q}),(\ref{annihilate varphi}),(\ref{fermfld corr bosfld1}) as $\psi_k=J_k$. Using this set of special solutions, the \textbf{loop equation} in the form of operators can be directly deduced.
\begin{equation}\label{eq:loop}
\begin{split}
  & \normord\displaystyle{\psi(z)\psi^*(w)}\normord\\
&=  \psi(z)\psi^*(w)-\frac{w}{z-w}\\
&= (\normord\displaystyle{e^{\phi(z)}}\normord\normord\displaystyle{e^{-\phi(w)}} \normord \displaystyle{w})-\frac{w}{z-w}\\
&=  \normord\displaystyle{e^{\phi(z)-\phi(w)}}\normord\displaystyle{\frac{w}{z-w}- \frac{w}{z-w}}\\
&=  \frac{w}{z-w}(\normord\displaystyle{e^{\phi(z)-\phi(w)}}\normord\displaystyle{-1}).
\end{split}
\end{equation}
The loop equation has been proven in another way in \cite{ACEH}. In \cite{AAEnumerative}, the loop equation can be used to derive the $W_{1+\infty}$ algebra.
\subsection{2D Toda $\tau$ functions and weighted Hurwitz numbers}
~\par
The notation and proofs of the details in this section can be found in \cite{HR,PJGen,PJ2D}.\par
For any group-like elements $G=\exp \Bigl (\sum_{i, k \in {\mathbb{z} }}b_{ik}\psi^*_i \psi_k\Bigr )$, it is well known that the KP $\tau$ functions is
\begin{equation}\label{KP tau}
\tau(\bt)=\lvs e^{J_+ ({\mathbf t})}G\rvs,
\end{equation}
where $\bt=\{t_1,t_2,\cdots\}$.\par
Using the generalized Wick theorem and the boson-fermion correspondence, S. Kharchev\cite{Ka} found the determinant representation of the KP $\tau$ functions:
\begin{equation}\label{KP det}
\begin{split}
\tau(\bt)= \frac{\lvs G\rvs}{\Delta(\nu)\nu_1\cdots\nu_N}\mathop{det} \limits_{i,j=1\cdots N}\Big(\frac{\lvs\psi_{-i}\psi^*(\nu_j)G\rvs}{\lvs G\rvs}\Big),
\end{split}
\end{equation}
where $\Delta(\nu)=\prod\limits_{1\leq i<j\leq N}(\nu_j-\nu_i)$ is the Vandermonde determinant and $t_k=\frac{1}{k}\sum_{i=1}^{N}\nu_i^{-k}$ is the Miwa parameters\cite{Miwa}.\par
The 2D Toda $\tau$ functions\cite{UT,Takebe} are
\begin{equation}\label{2DTL tau}
\tau^{2D} (n,{\mathbf t} , {\mathbf s})=
\lvsn e^{J_+ ({\mathbf t})}Ge^{J_- ({\mathbf s})}\rvsn,
\end{equation}
where $\bt=\{t_1,t_2,\cdots\}$ and $\mathbf{s}=\{s_1,s_2,\cdots\}$. In the $\bs=0$ case, the 2D Toda $\tau$ functions degenerate to the KP $\tau$ functions. Note that in\cite{AAFree}, the 2D Toda $\tau$ functions are defined as $\lvsn e^{J_+ ({\mathbf t})}Ge^{-J_- ({\mathbf s}) }\rvsn$, so the same formulas in this article differ a little from those in\cite{AAFree}. However, the present formulas compared with those of \cite{ACEH,ACEH2,HB,HO,PJGen,PJ2D} are consistent. \par
For any partition $\lambda$ and integer $n$, we have
\begin{equation}\label{schur1}
\displaystyle{
e^{J_-({\mathbf t})}\rvsn = \sum_{\lambda} (-1)^{b(\lambda )}s_{\lambda}({\mathbf t})\left |\lambda , n\rbr},
\end{equation}
\begin{equation}\label{schur2}
\displaystyle{
\lvsn e^{J_+({\mathbf t})}=\sum_{\lambda} (-1)^{b(\lambda )}s_{\lambda}({\mathbf t})\lbr \lambda , n \right |},
\end{equation}
where $s_{\lambda}({\mathbf t})$ is the Schur function\cite{Mac} and $b(\lambda )=\sum_{i=1}^{d(\lambda )}(\beta_i +1)$.\par
For partitions $\{\mu^{(i )}\}_{i=1, \dots, k}$ with the same weight, the Hurwitz numbers $H(\mu^{(1)},\mu^{(2)},\cdots,\mu^{(k)})$ are defined as the number of distinct factorizations of the identity element $I_N \in \mathbf{S}_N$ in the symmetric group on $N$ elements divided by the normalization factor $N!$. $I_N=h_1h_2\cdots h_k,\,cyc(h_i)=\mu^{(i)}$ where the conjugate class $cyc(\mu^{(i)})\in \mathbf{S}_N$ whose cycle lengths are equal to the partitions of $\mu^{(i)}$\cite{Frobenius1,Frobenius2,Schur}. Usually, we use the Frobenius-Schur formula:
\begin{equation}\label{Hurwitz}
H(\mu^{(1)}, \dots , \mu^{(k)}) = \sum_{|\lambda|=N}h(\lambda)^{k-2} \prod_{i=1}^k {\frac{\chi_\lambda(\mu^{(i)})}{Z_{\mu^{(i)}}}}.
\end{equation}
The calculation of the Hurwitz numbers gives an explicit expression of the Hurwitz numbers with hook lengths $h(\lambda)$, $Z_\mu=\prod_i m_i! i^{m_i}$ and the $\mathbf{S}_N$ characters $\chi_\lambda(\mu^{(i)})$ corresponding to the irreducible representations with symmetry class $\lambda$.\par
The weighted Hurwitz numbers $H^d_G(\mu,\nu)$ \cite{PJGen,HO} with the weight generating function $G(z)$ are
\begin{align}
G(z):= & \prod\limits_{k=1}^{\infty}(1+c_kz)=\sum_{k=0}^\infty G_k z^k
\end{align}
and $d\in \mathbb{N}^+$; two partitions $\mu,\nu$ with the same weight are
\begin{equation}\label{def:H_G}
\begin{split}
H^d_G(\mu,\nu):=           & \sum_{k=0}^\infty \sideset{}{'}\sum_{\substack{\mu^{(1)}\cdots\mu^{(k)},\\ |\mu^{(1)}|=\cdots=|\mu^{(k)}|=|\mu|\\ \sum_{i=1}^{k}l^*(\mu^{(i)})=d}}W_G(\mu^{(1)},\mu^{(2)},\cdots,\mu^{(k)}) H(\mu^{(1)},\mu^{(2)},\cdots,\mu^{(k)},\mu,\nu),
 \end{split}
\end{equation}
where $\sum'$ refers to the sum for all $ \mu^{(1)}\cdots\mu^{(k)}, \,|\mu^{(1)}|=\cdots=|\mu^{(k)}|=|\mu|,\,$ $\sum_{i=1}^{k}l^*(\mu^{(i)})=d$ except $\mu^{(i)}=(1^N)$, and the $W_G(\mu^{(1)},\mu^{(2)},\cdots,\mu^{(k)})$ are the weights of $H^d_G(\mu,\nu)$ related to the weight generating function $G(z)$:
\begin{equation}
\begin{split}
W_G(\mu^{(1)},\mu^{(2)},\cdots,\mu^{(k)})= & \frac{1}{k!}\sum_{\sigma\in S_k}\sum_{1\leq i_1<i_2<\cdots<i_k}c_{i_{\sigma(1)}}^{l^*(\mu^{(1)})}c_{i_{\sigma(2)}}^{l^*(\mu^{(2)})}\cdots c_{i_{\sigma(k)}}^{l^*(\mu^{(k)})},
\end{split}
\end{equation}
where $l^*(\mu):=|\mu|-l(\mu)$ is the co-length of the partition $\mu$, and the $c_k$ are the parameters in the weight generating function $G(z)$. $H^d_G(\mu,\nu)$ vanishes, if $\mu,\,\nu$ have different weights. The weighted Hurwitz numbers also have an equivalent combinatorial definition\cite{HR}. As for $d = 0$, the weighted Hurwitz numbers $H^d_G(\mu,\nu)$ degenerate to the Hurwitz numbers $H(\mu,\nu)$, no mater what the weight generating function is.\par
A parametric family $\tau^{2D}_{\hat{C}_\rho}(n,\bt,\mathbf{s})$ \cite{PJ2D,PJGen,HO} associated to the group-like elements $\hat{C}_\rho=e^{\sum_{j\in\mathbb{Z}}T_j\normord\displaystyle{\psi_j\psi_j^*}\normord}$ with $\rho=\{\rho_i\}_{i=-\infty}^\infty$, $\rho_i=e^{T_i}$, $\rho_{0}=1$, $r_i=e^{T_i-T_{i-1}}$, $r_k=G(\beta k)$,
\begin{equation}\label{C tau}
\tau^{2D}_{\hat{C}_\rho} (n,{\mathbf t} , {\mathbf s})=
\lvsn e^{J_+ ({\mathbf t})}\hat{C}_\rho e^{J_- ({\mathbf s})}\rvsn,
\end{equation}
are the generating functions of the weighted Hurwitz numbers and weighted paths in the Cayley graph of the symmetric group $\mathbf{S}_n$. We abbreviate $\tau^{2D}_{\hat{C}_\rho} (0,{\mathbf t} , {\mathbf s})$ as $\tau^{2D}_{\hat{C}_\rho}(\bt,\mathbf{s})$.
\begin{equation}\label{tau Hurwitz}
\tau^{2D}_{\hat{C}_\rho}(\bt,\mathbf{s})=\sum_{d=0}^\infty \beta^d \sum_{\substack{\mu,\nu\\|\mu|=|\nu|}}H^d_G(\mu,\nu)p_\mu(\bt)p_\nu(\mathbf{s}),
\end{equation}
in which $\bt$, $\mathbf{s}$ are the Miwa parameters\cite{Miwa}, and $\{\mu\}$, $\{\nu\}$ are called the Miwa variables
\begin{equation}\label{def:Miwa}
t_k=\frac{1}{k}\sum_{i=1}^{N}\nu_i^{-k},\qquad\qquad s_l=\frac{1}{l}\sum_{j=1}^{M}\mu_j^{-l}.
\end{equation}
Simultaneously,
\begin{equation}\label{Cayley}
H^d_G(\mu,\nu)=F^d_G(\mu,\nu)=\frac{1}{|\nu|!}\sum\limits_{\lambda,|\lambda|=d} G_\lambda m^\lambda_{\mu\nu},
\end{equation}
where $m^\lambda_{\mu\nu}$ is the number of monotonic $\lambda$ signature paths in the Cayley graph of $\mathbf{S}_n,\,|\mu|=|\nu|=n$ from $cyc(\mu)$ to $cyc(\nu)$.\par
To save space we will not list the specific proof here, but an intermediate step in the proof process will useful later:
\begin{equation}\label{C_rho}
\begin{split}
\hat{C}_\rho\left|\lambda,N\rbr =r_\lambda(N)\left|\lambda,N\rbr,
 \end{split}
\end{equation}
where $r_\lambda(N)=r_0(N)\prod\limits_{(i,j)}r_{N+j-i}$ and $r_i=e^{T_i-T_{i-1}}$ with
\begin{equation}
r_0(N)=
\left\{
\begin{array}{r}
\prod\limits_{j=0}^{N-1}e^{T_j}\qquad N>0,\\
1                       \qquad\qquad N=0,\\
\prod\limits_{j=N}^{-1}e^{-T_j}\qquad N<0.
 \end{array}
 \right.
\end{equation}
Furthermore, the 2D Toda $\tau$ functions could be written as double Schur function expansions\cite{HO}:
\begin{equation}\label{H_Schur}
\tau^{2D}_{\hat{C}_\rho}(n,\bt,\mathbf{s})= \sum_{\lambda}r_\lambda(n)s_\lambda(\bt)s_{\lambda}(\mathbf{s}).
\end{equation}
There is another parametric family $\tau^{2D}_{\hat{C}_\rho}(\gamma,n,\bt,\mathbf{s}) =\sum\limits_{N=0}^\infty\gamma^{N}\sum\limits_{\lambda,|\lambda|=N} r_\lambda(n)s_\lambda(\bt)s_{\lambda}(\mathbf{s})$. When $\gamma=1$, $\tau^{2D}_{\hat{C}_\rho}(\gamma,n,\bt,\mathbf{s}) =\tau^{2D}_{\hat{C}_\rho}(n,\bt,\mathbf{s})$.
\section{Determinant representation  of 2D Toda $\tau$ functions}
~\par
In the 1990s, S. Kharchev\cite{Ka} used the generalized Wick theorem and the boson-fermion correspondence\cite{Ka,DLM} to get the determinant of the KP(Kadomtsev-Petviashvili)$\tau$ function. We generalize that result to the case of the 2D Toda $\tau$ functions in this section.\par
By the correspondence between the Bosonic field and the Fermion field, we have
\begin{equation}\label{right 2D decom}
\begin{split}
  & \psi(\mu_1^{-1})\cdots\psi(\mu_M^{-1})\left|-M\rbr\\
= & \normord\displaystyle{e^{\phi(\mu_1^{-1})}}\normord \cdots\normord\displaystyle{e^{\phi(\mu_M^{-1})}}\normord\displaystyle{\left|-M \rbr}\\
= & \prod\limits_{i<j}(\mu_i^{-1}-\mu_j^{-1})e^{\sum\limits_{k=1}^{\infty} \frac{J_{-k}}{k}(\mu_1^{-k}+\cdots+\mu_M^{-k})}e^{MP}...
   \mu_1^{-Q}\cdots\mu_{M}^{-Q}e^{\sum\limits_{k=1}^{\infty} -\frac{J_{k}}{k}(\mu_1^{k}+\cdots+\mu_M^{k})}\displaystyle{\left|-M\rbr}\\
= & \Delta(\mu)\mu_1\cdots\mu_M\hat{\gamma}_-(\mathbf{s})\rvs,
\end{split}
\end{equation}
where $\Delta(\mu)=\prod\limits_{1\leq i<j\leq M}(\mu_j-\mu_i)$ is the Vandermonde determinant and $s_l=\frac{1}{l}\sum\limits_{j=1}^{M}\mu_j^{-l}$. Similarly, we have
\begin{equation}\label{left 2D decom}
\begin{split}
 &\lbr-N\right|\psi^*(\nu_N)\cdots\psi^*(\nu_1)=\Delta(\nu) \nu_1\cdots\nu_N\lvs\hat{\gamma}_+(\mathbf{t}).
\end{split}
\end{equation}
Observing that in (\ref{def:Miwa}), if we add $\nu_{N+1}=\infty$ after the sequence of Miwa variables $\left\{\mathbf{\nu}\right\}$, the new Miwa variables $\left\{\mathbf{\nu}'\right\}=\left\{\mathbf{\nu_1,\cdots,\nu_{N},\nu_{N+1}}\right\}$ also correspond to $\mathbf{t}$, but it changes the form of (\ref{left 2D decom}).
\begin{equation}\label{eq:2D infty}
\begin{split}
 \lbr-N-1\right|\psi^*(\nu_{N+1})\psi^*(\nu_N)\cdots\psi^*(\nu_1)
=\Delta(\nu)\prod\limits_{j=1}^{N}(\nu_{N+1}-\nu_j)\nu_1\cdots\nu_N\nu_{N+1} \lvs\hat{\gamma}_+(\mathbf{t}),
\end{split}
\end{equation}
where $\Delta(\nu)$ is the Vandermonde determinant of $\left\{\nu\right\}$, $\Delta(\nu)=\prod\limits_{1\leq i<j\leq N}(\nu_j-\nu_i)$.
Even though the form of (\ref{eq:2D infty}) corresponding to $\left\{\mathbf{\nu}'\right\}$ and (\ref{left 2D decom}) corresponding to $\left\{\mathbf{\nu}\right\}$ are different, both of them are equivalent.
For $\nu_{N+1}=\infty$, the left side of (\ref{eq:2D infty}) is
\begin{equation}
\begin{split}
  & \mathop{lim}_{\nu_{N+1}\rightarrow\infty}\lbr-N-1\right|\psi^*(\nu_{N+1}) \psi^*(\nu_N)\cdots\psi^*(\nu_1)\\
&~=  \mathop{lim}_{\nu_{N+1}\rightarrow\infty}\lbr-N-1\right|\nu_{N+1} \normord\displaystyle{e^{-\phi(\nu_{N+1})}}\normord\psi^*(\nu_N)\cdots\psi^*(\nu_1)\\
&~= \mathop{lim}_{\nu_{N+1}\rightarrow\infty}\nu_{N+1}\lbr -N\right|\nu_{N+1}^{-Q} e^{\sum_{k=1}^{\infty}\frac{J_{k}}{k}\nu_{N+1}^{-k}}\psi^*(\nu_N)\cdots\psi^*(\nu_1)\\
&~= \mathop{lim}_{\nu_{N+1}\rightarrow\infty}\nu_{N+1}^{N+1}\lbr -N\right|\psi^*(\nu_N)\cdots\psi^*(\nu_1),\\
\end{split}
\end{equation}
while the right side of (\ref{eq:2D infty}) is
\begin{equation}
\begin{split}
  & \mathop{lim}_{\nu_{N+1}\rightarrow\infty}\Delta(\nu)\prod\limits_{j=1}^{N} (\nu_{N+1}-\nu_j)\nu_1\cdots\nu_N \nu_{N+1}\lvs\hat{\gamma}_+(\mathbf{t})\\
&~= \mathop{lim}_{\nu_{N+1}\rightarrow\infty}\nu_{N+1}^{N+1} \Delta(\nu)\nu_1\cdots\nu_N\lvs\hat{\gamma}_+(\mathbf{t}).
\end{split}
\end{equation}
Dividing both sides of (\ref{eq:2D infty}) by $\nu_{N+1}^{N+1} $, we get (\ref{left 2D decom}). When the two Miwa variables $\{\mu\}$, $\{\nu\}$ have different dimensions, which is likely to happen as $\{\mu\}$ and $\{\nu\}$ are independent, we could add some $\infty$ term after the Miwa variables with the smaller dimension to make $\{\mu\}$, $\{\nu\}$ have the same dimensions. We only talk about the situation where $\{\mu\}$, $\{\nu\}$ have the same dimension N below.
Considering (\ref{right 2D decom}) and (\ref{left 2D decom}), we have\\
\begin{equation}\label{2D fermion}
\begin{split}
\tau^{2D}(\bt,\mathbf{s})= & \lvs \hat{\gamma}_+({\mathbf t})G\hat{\gamma}_-({\mathbf s})\rvs\\
= & \frac{1}{\Delta(\nu)\Delta(\mu)\nu_1\cdots\nu_N\mu_1\cdots\mu_N}\lbr-N\right |\psi^*(\nu_N)\cdots\psi^*(\nu_1)\\
  & G\psi(\mu_1^{-1})\cdots\psi(\mu_N^{-1})\left|-N\rbr.
\end{split}
\end{equation}
In the $\mathbf{s}=0$ case, which means $\nu_i\rightarrow\infty,\,\forall i$, the right side of (\ref{2D fermion}) degenerates into $\frac{1}{\Delta(\nu)\nu_1\cdots\nu_N} $ $\lbr-N\right|\psi^*(\nu_N) \cdots\psi^*(\nu_1)G\rvs$ corresponding to the KP $\tau$ functions $\tau(\bt)$ in (\ref{KP det}). \par
From the corollary of the generalized Wick theorem, let $G=G'=1$ which satisfies the BBC and is a group-like element. We can get a determinant representation of the 2D Toda $\tau$ functions:
\begin{equation}\label{2DTL det}
\begin{split}
\tau^{2D}(\bt,\mathbf{s})= & \lvs \hat{\gamma}_+({\mathbf t})G\hat{\gamma}_-({\mathbf s})\rvs\\
= & \frac{\lbr-N\right| G\left|-N\rbr}{\Delta(\nu)\Delta(\mu)\nu_1\cdots\nu_N\mu_1\cdots\mu_N} \mathop{det}\limits_{i,j=1\cdots N}\Big(\frac{\lbr-N\right|\psi^*(\nu_j)G\psi(\mu_i^{-1})\left|-N\rbr}{\lbr-N\right| G\left|-N\rbr}\Big).
\end{split}
\end{equation}
Such a determinant representation of the 2D Toda $\tau$ functions can be regarded as generalized KP $\tau$ functions, while in the $\mathbf{s}=0$ case, the 2D Toda $\tau$ functions degenerate KP $\tau$ functions, we have to prove that the determinant representation of the 2D Toda $\tau$ functions degenerates to a determinant representation of the KP $\tau$ functions in the case of $\mathbf{s}=0$.\par
We get (\ref{2DTL det}) and (\ref{KP det}) by using different forms of the generalized Wick theorem. At first, we need to use the generalized Wick theorem to get an equivalent form of (\ref{KP det}).\\
\begin{equation}\label{KP det2}
\begin{split}
\tau(\bt)=  & \lvs \hat{\gamma}_+({\mathbf t})G\rvs\\
= & \frac{1}{\Delta(\nu)\nu_1\cdots\nu_N}\lbr-N\right |\psi^*(\nu_N)\cdots\psi^*(\nu_1)G\psi_{-1}\cdots\psi_{-N}\left|-N\rbr\\
= & \frac{\lbr-N\right|G\left|-N\rbr}{\Delta(\nu)\nu_1\cdots\nu_N}\mathop{det} \limits_{i,j=1\cdots N}\Big(\frac{\lbr-N\right|\psi^*(\nu_j)G\psi_{-i} \left|-N\rbr}{\lbr-N\right| G\left|-N\rbr}\Big).
\end{split}
\end{equation}
Now, we only need to prove that (\ref{2DTL det}) degenerates to (\ref{KP det2}) when $\mathbf{s}=0$.\par
$\mathbf{s}=0$ means $\mu_i\rightarrow\infty,\quad\forall i$,
\begin{align}\label{2D degen KP}
  & \mathop{lim}_{\mu\rightarrow\infty}\frac{\lbr-N\right| G\left|-N\rbr} {\Delta(\nu)\Delta(\mu)\nu_1\cdots\nu_N\mu_1\cdots\mu_N}\mathop{det}\limits_{i,j=1\cdots N}\Big(\frac{\lbr-N\right|\psi^*(\nu_j)G\psi(\mu_i^{-1})\left|-N\rbr}{\lbr-N\right| G\left|-N\rbr}\Big)\\
= & \mathop{lim}_{\mu\rightarrow\infty}\frac{\lbr-N\right| G\left|-N\rbr} {\Delta(\nu)\Delta(\mu)\nu_1\cdots\nu_N\mu_1\cdots\mu_N}\mathop{det}\limits_{i,j=1\cdots N}\Big(\frac{\lbr-N\right|\psi^*(\nu_j)G\sum\limits_{i=1}^{N}\psi_{-i}\mu_{k}^{i} \left|-N\rbr}{\lbr-N\right| G\left|-N\rbr}\Big)\notag\\
= & \mathop{lim}_{\mu\rightarrow\infty}\frac{\lbr-N\right|G\left|-N\rbr} {\Delta(\nu)\Delta(\mu) \nu_1\cdots\nu_N\mu_1\cdots\mu_N} \mathop{det}\limits_{k,i=1\cdots N}(\mu_{k}^{i})\mathop{det}\limits_{i,j=1\cdots N}\Big(\frac{\lbr-N\right|\psi^*(\nu_j)G\psi_{-i}\left|-N\rbr}{\lbr-N\right| G\left|-N\rbr}\Big)\notag\\
= & \frac{\lbr-N\right|G\left|-N\rbr}{\Delta(\nu)\nu_1\cdots\nu_N}\mathop{det} \limits_{i,j=1\cdots N}\Big(\frac{\lbr-N\right|\psi^*(\nu_j)G\psi_{-i} \left|-N\rbr}{\lbr-N\right| G\left|-N\rbr}\Big).\notag
\end{align}
The second equal in (\ref{2D degen KP}) is sure, as $\mu_i^n=0$, $\forall n<0$ for $\mu_i=\infty$ and
\begin{equation}
\begin{split}
\psi(\mu_i^{-1})\left|-N\rbr=\sum\limits_{i=-\infty}^{\infty}\psi_{-i}\mu_{k}^{i} \left|-N\rbr =\sum\limits_{i=-\infty}^{N}\psi_{-i}\mu_{k}^{i} \left|-N\rbr =\sum\limits_{i=1}^{N}\psi_{-i}\mu_{k}^{i} \left|-N\rbr.
\end{split}
\end{equation}
Finally, the invertible matrix variable $t_k=\frac{1}{k}Tr(Z^{-k})$, where Z is a diagonal matrix whose diagonal elements are the Miwa parameters; it could also be written as the determinant representation of the KP $\tau$ functions \cite{AAEnumerative}. This expression is similar to the KP $\tau$ functions in terms of the Miwa parameters, and it has a very important application in the Generalized Kontsevich Matrix Model (GKMM)\cite{Ko}.
\begin{equation}\label{eq:KP matrix}
\begin{split}
\tau([Z^{-1}])= & \lvs e^{J_{+}([z_1^{-1}])+\cdots+J_{+}([z_N^{-1}])}G\rvs\\
              = & \frac{\lvs \psi_{-1}\cdots\psi_{-N}\psi^*(z_N)\cdots\psi^*(z_1)G\rvs}{z_1\cdots z_N\Delta(z)}\\
              = & \lvs G\rvs\frac{\mathop{det}_{i,j=1\cdots N}f_i^*(z_j)}{\Delta(z)},
\end{split}
\end{equation}
where $f_i^*(z)=z^{-1}\frac{\lvs \psi_{-i}\psi^*(z)G\rvs}{\lvs G\rvs}$.\par
For the 2D Toda $\tau$ functions
\begin{equation}\label{eq:2D matrix}
\begin{split}
\tau^{2D}([Z^{-1}],[W^{-1}])= & \lvs e^{J_{+}([z_1^{-1}])+\cdots+J_{+}([z_N^{-1}])}Ge^{J_{+}([w_N^{-1}])+\cdots+J_{+}([w_1^{-1}])}\rvs\\
              = & \frac{\lbr-N\right|\psi^*(z_N)\cdots\psi^*(z_1)G\psi(w_1)\cdots\psi(w_N)\left|-N\rbr}{z_1\cdots z_Nw_1\cdots w_N\Delta(z)\Delta(w)}\\            = & \lbr-N\right| G\left|-N\rbr\frac{\mathop{det}_{i,j=1\cdots N}f^*(z_j,w_i)}{\Delta(z)\Delta(w)},\\
\end{split}
\end{equation}
where $f^*(z,w)=z^{-1}w^{-1}\frac{\lbr-N\right|\psi^*(z)G\psi(w)\left|-N\rbr} {\lbr-N\right|G\left|-N\rbr}$. From (\ref{KP det2}) and (\ref{2D degen KP}), (\ref{eq:2D matrix}) degenerates to (\ref{eq:KP matrix}) as $w_i\rightarrow\infty$.
\section{Determinant representation of the Weighted Hurwitz numbers}
~\par
We got the determinant representation of the 2D Toda $\tau$ functions in the last section. For a parametric family $\tau^{2D}_{\hat{C}_\rho}(n,\bt,\mathbf{s})$ (\ref{C tau}) associated to the group-like elements $\hat{C}_\rho=e^{\sum_{j\in\mathbb{Z}}T_j\normord\displaystyle{\psi_j\psi_j^*}\normord}$ with $\rho=\{\rho_i\}_{i=-\infty}^\infty$, $\rho_i=e^{T_i}$, $r_i=e^{T_i-T_{i-1}}$, $r_k=G(\beta k)$,
\begin{equation}\nonumber
\tau^{2D}_{\hat{C}_\rho} (n,{\mathbf t} , {\mathbf s})=
\lvsn e^{J_+ ({\mathbf t})}\hat{C}_\rho e^{J_- ({\mathbf s})}\rvsn
\end{equation}
has the determinant representation
\begin{equation}\label{2D C det}
\begin{split}
\tau^{2D}_{\hat{C}_\rho}(\bt,\mathbf{s})= & \lvs e^{J_+ ({\mathbf t})}\hat{C}_\rho e^{J_- ({\mathbf s})}\rvs\\
= & \frac{\lbr-N\right| \hat{C}_\rho \left|-N\rbr}{\Delta(\nu)\Delta(\mu)\nu_1\cdots\nu_N\mu_1\cdots\mu_N}\mathop{det}\limits_{i,j=1\cdots N}\Big(\frac{\lbr-N\right|\psi^*(\nu_j)\hat{C}_\rho \psi(\mu_i^{-1})\left|-N\rbr} {\lbr-N\right| \hat{C}_\rho \left|-N\rbr}\Big)
\end{split}
\end{equation}
with the Miwa parameters
\begin{equation}
t_k=\frac{1}{k}\sum_{i=1}^{N}\nu_i^{-k},\qquad\qquad s_l=\frac{1}{l}\sum_{j=1}^{M}\mu_j^{-l}.
\end{equation}\par
From (\ref{C_rho}), we find
\begin{equation}
\begin{split}
\tau^{2D}_{\hat{C}_\rho}(\bt,\mathbf{s}) & =\frac{\lbr-N\left|\hat{C}_\rho\right|-N\rbr}{\Delta(\nu)\Delta(\mu)\nu_1\cdots\nu_N \mu_1\cdots\mu_N}\mathop{det}\limits_{i,j=1\cdots N}\Big(\frac{\lbr-N\left|\psi^*(\nu_j)\hat{C}_\rho\psi(\mu_i^{-1}) \right|-N\rbr}{\lbr-N\right|\hat{C}_\rho\left|-N\rbr}\Big)\\
                                        %& =\frac{r_0(-N)}{\Delta(\nu)\Delta(\mu)\nu_1\cdots\nu_N\mu_1\cdots\mu_N} \mathop{det}\limits_{i,j=1\cdots N}\Big(\frac{\lbr-N\right|\sum_{k\in\mathbb{Z}}\psi^*_k \nu_j^{-k} \hat{C}_\rho\sum_{l\in\mathbb{Z}}\psi_l \mu_i^{-l}\left|-N\rbr} {r_0(-N)}\Big)\\
                                        & =\frac{r_0(-N)}{\Delta(\nu)\Delta(\mu)\nu_1\cdots\nu_N \mu_1\cdots\mu_N}\mathop{det}\limits_{i,j=1\cdots N}\Big(\frac{\lbr-N\right|\sum_{k\in\mathbb{Z}}\psi^*_k \nu_j^{-k} \hat{C}_\rho\sum_{l\in\mathbb{Z}}\psi_l \mu_i^{-l}\left|-N\rbr} {r_0(-N)}\Big)\\
                                        & =\frac{r_0(-N)}{\Delta(\nu)\Delta(\mu)} \mathop{det}\limits_{i,j=1\cdots N}\Big(\sum_{l=-N}^\infty \nu_j^{-l-1}\mu_i^{-l-1}\rho_l\Big).
\end{split}
\end{equation}
The last equality holds as
\begin{align}
  & \lbr-N\right|\sum_{k\in\mathbb{Z}}\psi^*_k \nu_j^{-k} \hat{C}_\rho\sum_{l\in\mathbb{Z}}\psi_l \mu_i^{-l}\left|-N\rbr\\
= & \sum_{k,l=-N}^\infty\lbr-N\right|\psi^*_k  \hat{C}_\rho\psi_l \left|-N\rbr\nu_j^{-k}\mu_i^{-l}\notag\\
= & \sum_{k,l=-N}^\infty\lbr-N\right|\psi^*_k  \hat{C}_\rho(-1)\left|(l+N),-N+1\rbr\nu_j^{-k}\mu_i^{-l},
= & \sum_{l=-N}^\infty \nu_j^{-l}\mu_i^{-l} r_0(-N+1)\frac{\rho_l}{\rho_{-N}},\notag
\end{align}
where $(l+N)=(l+N,0,0,\cdots)$.\par
Recalling (\ref{tau Hurwitz}) \cite{PJGen}, we have
\begin{equation}\label{tep5}
\sum_{d=0}^\infty \beta^d \sum_{\substack{\omega,\sigma\\|\omega|=|\sigma|}}H^d_G(\omega,\sigma)p_\omega(\bt)p_\sigma(\mathbf{s})
=\frac{r_0(-N)}{\Delta(\nu)\Delta(\mu)}\mathop{det}\limits_{i,j=1\cdots N}\Big(\sum_{l=-N}^\infty \nu_j^{-l-1}\mu_i^{-l-1}\rho_l\Big).
\end{equation}
Notice that the left side of (\ref{tep5}) also contains $N$ explicitly appearing in $t_k=\frac{1}{k}\sum\limits_{i=1}^{N}\nu_i^{-k},\,s_l =\frac{1}{l}\sum\limits_{j=1}^{N}\mu_j^{-l}$ and (\ref{tep5}) are not the same for different $N$.
We could use (\ref{tep5}) through certain determinant transformations rather than the operator approach to obtain the double Schur function expansions of the 2D Toda $\tau$ functions $\tau^{2D}_{\hat{C}_\rho}(\bt,\mathbf{s})$ (\ref{H_Schur})\cite{HO}, which guarantees the correctness of the determinant representation of the 2D Toda $\tau$ functions (\ref{2DTL det}) and the determinant representation of the weighted Hurwitz numbers (\ref{tep5}).
\begin{align}\label{tep31}
 \sum &_{d=0}^\infty \beta^d \sum_{\substack{\omega,\sigma\\|\omega|=|\sigma|}}H^d_G(\omega,\sigma) p_\omega(\bt)p_\sigma(\mathbf{s})\\
= & \frac{r_0(-n)}{\Delta(\nu)\Delta(\mu)}\mathop{det}\limits_{i,j=1\cdots n}\Big(\sum_{l=-n}^\infty \nu_j^{-l-1}\mu_i^{-l-1}\rho_l\Big)\notag\\
= & \frac{r_0(-n)}{\Delta(\nu)\Delta(\mu)}\sum_{\substack{k_1,\cdots,k_n=-\infty\\k_1>\cdots>k_n}}^{n-1}\sum_{\sigma\in \mathbf{S}_n}sgn(\sigma)\notag\\
  & \left|\begin{array}{cccc}
\nu_1^{k_{\sigma(1)}}\mu_1^{k_{\sigma(1)}}\rho_{-k_{\sigma(1)}-1} & \nu_2^{k_{\sigma(2)}}\mu_1^{k_{\sigma(2)}}\rho_{-k_{\sigma(2)}-1} & \cdots & \nu_n^{k_{\sigma(1)}}\mu_1^{k_{\sigma(n)}}\rho_{-k_{\sigma(n)}-1}\\
\nu_1^{k_{\sigma(1)}}\mu_2^{k_{\sigma(1)}}\rho_{-k_{\sigma(1)}-1} & \nu_2^{k_{\sigma(2)}}\mu_2^{k_{\sigma(2)}}\rho_{-k_{\sigma(2)}-1} & \cdots & \nu_n^{k_{\sigma(1)}}\mu_2^{k_{\sigma(n)}}\rho_{-k_{\sigma(n)}-1}\\
\vdots & \vdots & \ddots & \vdots\\
\nu_1^{k_{\sigma(1)}}\mu_n^{k_{\sigma(1)}}\rho_{-k_{\sigma(1)}-1} & \nu_2^{k_{\sigma(2)}}\mu_n^{k_{\sigma(2)}}\rho_{-k_{\sigma(2)}-1} & \cdots & \nu_n^{k_{\sigma(1)}}\mu_n^{k_{\sigma(n)}}\rho_{-k_{\sigma(n)}-1}\\
\end{array}\right|\notag\\
= & \frac{r_0(-n)}{\Delta(\nu)\Delta(\mu)}\sum_{\substack{k_1,\cdots,k_n=-\infty\\k_1>\cdots>k_n}}^{n-1}\rho_{-k_{\sigma(1)}-1}\cdots\rho_{-k_{\sigma(n)}-1} \sum_{\sigma,\omega\in \mathbf{S}_n}sgn(\sigma)sgn(\omega)\notag\\
  & \nu_{1}^{k_\sigma(1)}\nu_{2}^{k_\sigma(2)}\cdots\nu_{n}^{k_\sigma(n)} \mu_{\omega(1)}^{k_\sigma(1)}\mu_{\omega(2)}^{k_\sigma(2)}\cdots \mu_{\omega(n)}^{k_\sigma(n)}\notag\\
= & \frac{r_0(-n)}{\Delta(\nu)\Delta(\mu)}\sum_{\substack{k_1,\cdots,k_n=-\infty\\ k_1>\cdots>k_n}}^{n-1}\rho_{-k_1-1}\cdots\rho_{-k_n-1}\sum_{\sigma,\omega\in \mathbf{S}_n}sgn(\sigma)sgn(\omega)\nu_{\sigma(1)}^{k_1}\nu_{\sigma(2)}^{k_2}\notag\\
  & \cdots\nu_{\sigma(n)}^{k_n} \mu_{\omega(1)}^{k_1}\mu_{\omega(2)}^{k_2}\cdots\mu_{\omega(n)}^{k_n}\notag\\
= & \sum_{\substack{k_1,\cdots,k_n=-\infty\\k_1>\cdots>k_n}}^{n-1} \frac{\rho_{-k_1-1}\cdots\rho_{-k_n-1}}{\rho_{-n}\cdots\rho_{-1}}\sum_{\sigma\in \mathbf{S}_n}sgn(\sigma) \frac{\nu_{\sigma(1)}^{k_1}\nu_{\sigma(2)}^{k_2}\cdots\nu_{\sigma(n)}^{k_n}} {\Delta(\nu)}\notag\\
  & \times\sum_{\omega\in \mathbf{S}_n}sgn(\omega) \frac{\mu_{\omega(1)}^{k_1} \mu_{\omega(2)}^{k_2}\cdots\mu_{\omega(n)}^{k_n}} {\Delta(\mu)}\notag\\
= & \sum_{\lambda}r_\lambda(0)s_\lambda(\bt)s_{\lambda}(\mathbf{s}).\notag
\end{align}
For convenience, we simplify the notation; for brevity let
\begin{equation}\label{eq:abbreviate}
\mu^\lambda=\sum_{\sigma\in \mathbf{S}_n}\mu_1^{\lambda_{\sigma(1)}}\cdots\mu_n^{\lambda_{\sigma(n)}}
\end{equation}
for all $\lambda,\,n\geq l(\lambda)$, when $l(\lambda)<n$, $\lambda_{i}=0,\,\forall i>l(\lambda)$. One can find the same notation in \cite{Mac}.\par
As $\bt$ and $\mathbf{s}$ are independent of each other, the coefficients of the same term $\mu^{-\lambda_1}\nu^{-\lambda_2},\,|\lambda_1|=|\lambda_2|$ on the two sides of (\ref{tep5}) are the same. We could find an infinite dimensional system of equations of weighted Hurwitz numbers by (\ref{tep5}). The 2D Toda $\tau$ functions are symmetric functions, we only need to discuss the coefficients of $\mu^{-\lambda_1}\nu^{-\lambda_2},\,|\lambda_1|=|\lambda_2|$ instead of each monomial of $\mu^{-\lambda_1}\nu^{-\lambda_2}$. And the coefficients of the two monomials are different, only if the two monomials belong to the same equivalence class under the action of the symmetry group.\par

Furthermore, if we collect all terms of the power $n$ on the two sides of (\ref{tep5}) together, we get a finite-dimensional equation system:
\begin{equation}\label{tep5-1}
\begin{split}
  & \sum_{d=0}^\infty \beta^d \sum_{\substack{\omega,\sigma\\|\omega|=|\sigma|=n}}H^d_G(\omega,\sigma)p_\omega(\bt)p_\sigma(\mathbf{s}) =Term_{-n}\Big\{\frac{r_0(-N)}{\Delta(\nu)\Delta(\mu)}\mathop{det}\limits_{i,j=1\cdots N}\Big(\sum_{l=-N}^\infty \nu_j^{-l-1}\mu_i^{-l-1}\rho_l\Big)\Big\},
\end{split}
\end{equation}
where $Term_{-n}\{f(\nu^{-1},\mu^{-1})=f(\nu_1^{-1},\cdots,\nu_N^{-1},\mu_1^{-1},\cdots,\mu_N^{-1})\}$ denotes the terms of ${\nu ^{-n}  \mu ^{-n}}$, which means the sum of the powers of $\{\mu_i\}_{i=1}^N$ or $\{\nu_j\}_{j=1}^N$ are $-n$, in the function $f(\nu^{-1},\mu^{-1})$ which is a polynomial with respect to $\{\nu_j^{-1}\}$ and $\{\mu_i^{-1}\}$. For $N\geq n$, there exist $p(n)^2$ equations in (\ref{tep5-1}) and the solutions of (\ref{tep5-1}) are the weighted Hurwitz numbers of the same dimension $|\omega|=|\sigma|=n$. Meanwhile, for the $n$ dimensional weighted Hurwitz numbers, its number is $p(n)^2$. The partition function $p(n)$ is the number of partitions of weight $n$\cite{Andrews}. That is to say the equation system is appropriate, and we could get the value of the $n$ dimensional weighted Hurwitz numbers by (\ref{tep5-1}).\par
For the $N<n$ case, (\ref{tep5-1}) is an underdetermined system of equations, such that the number of the power $-n$ term on the right side of (\ref{tep5}) is less than $p(n)^2$. This system of equations is not enough to determine the value of the $n$ dimensional weighted Hurwitz numbers $H^d_G(\omega,\sigma)$,\,$|\omega|=|\sigma|=n$, but we still could find certain interesting concise equations of the $n$ dimensional weighted Hurwitz numbers.

This brings a problem, as long as $N\geq n$, we could get the $n$ dimensional weighted Hurwitz numbers $H^d_G(\omega,\sigma)\,|\omega|=|\sigma|=n$ through (\ref{tep5}). Then for different $N$, are the obtained $n$ dimensional weighted Hurwitz numbers the same? \par

If $\mu_{N+1},\nu_{N+1}$ tends to negative infinity, the dimension of the Miwa parameters is still N+1, such as $\hat{t}_k=\frac{1}{k} \sum_{i=1}^{N+1}\nu_i^{-k},\,\hat{s}_l=\frac{1}{l}\sum_{j=1}^{N+1 }\mu_j^{-l}$. So (\ref{tep5}) becomes
\begin{equation}\label{tep_N+1}
\begin{split}
\sum_{d=0}^\infty \beta^d \sum_{\substack{\omega,\sigma\\|\omega|=|\sigma|}}H^d_G(\omega,\sigma) p_\omega(\hat{\bt})p_\sigma(\hat{\mathbf{s}}) =\frac{r_0(-N-1)}{\Delta(\nu)\Delta(\mu)}\mathop{det}\limits_{i,j=1\cdots N+1}\Big(\sum_{l=-N-1}^\infty \nu_j^{-l-1}\mu_i^{-l-1}\rho_l\Big).
\end{split}
\end{equation}
The right side of (\ref{tep_N+1}) is
\begin{align}\label{tep25}
  & \mathop{lim}_{\mu_{N+1},\nu_{N+1}\rightarrow-\infty}\frac{r_0(-(N+1))} {\Delta(\nu_1\cdots\nu_{N+1})\Delta(\mu_1\cdots\mu_{N+1})} \mathop{det}\limits_{i,j=1\cdots N+1}\Big(\sum\limits_{l=-N-1}^\infty \nu_j^{-l-1}\mu_i^{-l-1}\rho_l\Big)\\
= & \mathop{lim}_{\mu_{N+1},\nu_{N+1}\rightarrow-\infty}\frac{r_0(-N)} {\Delta(\nu_1\cdots\nu_{N})\Delta(\mu_1\cdots\mu_{N})} \frac{1}{\rho_{-N-1}\mu_{N+1}^N\nu_{N+1}^N} \notag\\
  & \times\left|\begin{array}{cccc}
\sum\limits\limits_{l=-N-1}^\infty \nu_1^{-l-1}\mu_1^{-l-1}\rho_l & \sum\limits_{l=-N-1}^\infty \nu_2^{-l-1}\mu_1^{-l-1}\rho_l & \cdots & \sum\limits_{l=-N-1}^\infty\nu_{N+1}^{-l-1}\mu_1^{-l-1}\rho_l\\
\sum\limits_{l=-N-1}^\infty \nu_1^{-l-1}\mu_2^{-l-1}\rho_l & \sum\limits_{l=-N-1}^\infty \nu_2^{-l-1}\mu_2^{-l-1}\rho_l & \cdots & \sum\limits_{l=-N-1}^\infty\nu_{N+1}^{-l-1}\mu_2^{-l-1}\rho_l\\
\vdots & \vdots & \ddots & \vdots\\
\sum\limits_{l=-N-1}^\infty \nu_1^{-l-1}\mu_{N+1}^{-l-1}\rho_l & \sum\limits_{l=-N-1}^\infty \nu_2^{-l-1}\mu_{N+1}^{-l-1}\rho_l & \cdots & \sum\limits_{l=-N-1}^\infty\nu_{N+1}^{-l-1}\mu_{N+1}^{-l-1}\rho_l
\end{array}\right|\notag\\
= & \mathop{lim}_{\mu_{N+1},\nu_{N+1}\rightarrow-\infty}\frac{r_0(-N)} {\Delta(\nu_1\cdots\nu_{N})\Delta(\mu_1\cdots\mu_{N})} \frac{1}{\rho_{-N-1}\mu_{N+1}^N\nu_{N+1}^N}\notag\\
  & \times\left|\begin{array}{cccc}
\sum\limits_{l=-N}^\infty \nu_1^{-l-1}\mu_1^{-l-1}\rho_l & \sum\limits_{l=-N}^\infty \nu_2^{-l-1}\mu_1^{-l-1}\rho_l & \cdots & \sum\limits_{l=-N}^\infty\nu_{N}^{-l-1}\mu_1^{-l-1}\rho_l\\
\sum\limits_{l=-N}^\infty \nu_1^{-l-1}\mu_2^{-l-1}\rho_l & \sum\limits_{l=-N}^\infty \nu_2^{-l-1}\mu_2^{-l-1}\rho_l & \cdots & \sum\limits_{l=-N}^\infty\nu_{N}^{-l-1}\mu_2^{-l-1}\rho_l\\
\vdots & \vdots & \ddots & \vdots\\
\sum\limits_{l=-N}^\infty \nu_1^{-l-1}\mu_{k}^{-l-1}\rho_l & \sum\limits_{l=-N}^\infty \nu_2^{-l-1}\mu_{k}^{-l-1}\rho_l & \cdots & \sum\limits_{l=-N}^\infty\nu_{N}^{-l-1}\mu_{k}^{-l-1}\rho_l
\end{array}\right|\notag\\
 & \times\rho_{-N-1}\mu_{N+1}^N\nu_{N+1}^N\notag\\
= & \frac{r_0(-N)}{\Delta(\nu_1\cdots\nu_{N})\Delta(\mu_1\cdots\mu_{N})}\mathop{det}\limits_{i,j=1\cdots N}\Big(\sum\limits_{l=-N}^\infty \nu_j^{-l-1}\mu_i^{-l-1}\rho_l\Big).\notag
\end{align}
Note that the determinant element summation starts from $l=-N$ for the second equal line, and the determinant element summation starts from $l=-N-1$ for the first equal line, for
\begin{equation}\label{tep24}
\begin{split}
  & \mathop{det}\limits_{i,j=1\cdots N+1}\Big(\sum_{l=-N-1}^\infty \nu_j^{-l-1}\mu_i^{-l-1}\rho_l\Big)\\
= & \sum_{\substack{l_1,\cdots,l_{N+1}=-N-1}}^{\infty}\left|\begin{array}{cccc}
\nu_{1}^{-l_1-1}\mu_{1}^{-l_1-1}\rho_{l_1} & \nu_{2}^{-l_2-1}\mu_{1}^{-l_2-1}\rho_{l_2} & \cdots & \nu_{N+1}^{-l_{N+1}-1}\mu_{1}^{-l_{N+1}-1}\rho_{l_{N+1}}\\
\nu_{1}^{-l_1-1}\mu_{2}^{-l_1-1}\rho_{l_1} & \nu_{2}^{-l_2-1}\mu_{2}^{-l_2-1}\rho_{l_2} & \cdots & \nu_{N+1}^{-l_{N+1}-1}\mu_{2}^{-l_{N+1}-1}\rho_{l_{N+1}}\\
\vdots & \vdots & \ddots & \vdots\\
\nu_{1}^{-l_1-1}\mu_{N+1}^{-l_1-1}\rho_{l_1} & \nu_{2}^{-l_2-1}\mu_{N+1}^{-l_2-1}\rho_{l_2} & \cdots & \nu_{N+1}^{-l_{N+1}-1}\mu_{k}^{-l_{N+1}-1}\rho_{l_{N+1}}\\
\end{array}\right|
\end{split}
\end{equation}
and $\{\mu_{N+1},\,\nu_{N+1}\}$ tend to negative infinity. If there is an $i\neq N+1$ having $l_i=-N-1$, then the $i$-th column of the determinant is proportional to the $(N+1)$-th column, that is to say the determinant is equal to 0.
Meanwhile, as well as $\mu_{N+1},\nu_{N+1}$ tending to negative infinity,
\begin{align}
  & \mathop{lim}_{\mu_{N+1},\nu_{N+1}\rightarrow-\infty}p_\omega(\hat{\bt})\\
&=~  \mathop{lim}_{\mu_{N+1},\nu_{N+1}\rightarrow-\infty}\big(\sum_{i=1}^{N+1} \mu_i^{-\omega_1}\big)\big(\sum_{i=1}^{N+1}\mu_i^{-\omega_1}\big) \cdots\big(\sum_{i=1}^{N+1}\mu_i^{-\omega_{l(\omega)}}\big)\notag\\
&=~ \big(\sum_{i=1}^{N}\mu_i^{-\omega_1}\big)\big(\sum_{i=1}^{N} \mu_i^{-\omega_1}\big) \cdots\big(\sum_{i=1}^{N}\mu_i^{-\omega_{l(\omega)}}\big)\notag\\
&=~ p_\omega(\bt)\notag.
\end{align}
Similarly, it is true for $p_\sigma(\hat{\mathbf{s}})$. \par
As well as $\mu_{N+1},\nu_{N+1}$ tending to negative infinity, (\ref{tep5}) for $N+1$ degenerates into (\ref{tep5}) for $N$. For the coefficients of the two monomials are the same as the two monomials belong to the same equivalence class under the action of the symmetry group, we can only compare the monomial of $\{\mu_1\cdots\mu_N\}^{-\lambda_1}\{\nu_1\cdots\nu_N\}^{-\lambda_2}, \,|\lambda_1|=|\lambda_2|=n\leq N$ rather than $\{\mu_1\cdots\mu_{N+1}\}^{-\lambda_1}\{\nu_1\cdots\nu_{N+1}\}^{-\lambda_2}, \,|\lambda_1|=|\lambda_2|=n\leq N$. We are sure that the two equation systems($N\geq n$) of the $n$ dimensional weighted Hurwitz numbers $H^d_G(\omega,\sigma),\,|\omega|=|\sigma|=n$ are the same. That is to say, we get the hierarchies of the weighted Hurwitz numbers for dimension $n$.\par
To simplify the calculation, in equation (\ref{tep5}) we use $N=n$ to calculate the $n$ dimensional weighted Hurwitz numbers $H^d_G(\omega,\sigma),\,|\omega|=|\sigma|=n$:
\begin{equation}\label{tep5-2}
\begin{split}
\sum_{d=0}^\infty \beta^d \sum_{\substack{\omega,\sigma\\|\omega|=|\sigma|=n}}H^d_G(\omega,\sigma)p_\omega(\bt)p_\sigma(\mathbf{s}) =Term_{-n}\Big\{\frac{r_0(-n)}{\Delta(\nu)\Delta(\mu)}\mathop{det}\limits_{i,j=1\cdots N}\Big(\sum_{l=-n}^\infty \nu_j^{-l-1}\mu_i^{-l-1}\rho_l\Big)\Big\}.
\end{split}
\end{equation}
\section{Weighted Hurwitz number and recursion formula}
~\par
In this section, we will calculate the value of the weighted Hurwitz numbers for the case where $n$ is $0,\,1,\,2,\,3$, and give a recursion formula for calculating the higher dimensional weighted Hurwitz number hierarchies.\par
First, the simplest case is $N=n=1$, which means $t_k=\frac{1}{k}\nu^{-k}\,,s_l=\frac{1}{l}\mu^{-l}$, then (\ref{tep5-2}) could be reduced to a rather concise form:
\begin{equation}
\sum_{d=0}^\infty \beta^d \sum_{n=0}^\infty\sum_{\substack{\omega,\sigma\\|\omega|=|\sigma|=n}}H^d_G(\omega,\sigma)(\nu\mu)^{-n}
=\sum_{l=-1}^\infty \nu^{-l-1}\mu^{-l-1}\frac{\rho_l}{\rho_{-1}}.
\end{equation}\par
For $\rho_{-1}=e^{T_{-1}}=G(0)=1$, we compare the coefficients of $(\nu\mu)^{-n}$ and find
\begin{equation}\label{tep9}
\sum_{d=0}^\infty \beta^d \sum_{\substack{\omega,\sigma\\|\omega|=|\sigma|=n}} H^d_G(\omega,\sigma)=\rho_{n-1}.
\end{equation}
We can directly get the value of the $0$ dimensional or $1$ dimensional weighted Hurwitz number
\begin{equation}
\sum_{d=0}^\infty \beta^d \sum_{\substack{\omega,\sigma\\|\omega|=|\sigma|=n}}H^d_G(\omega,\sigma)=1.
\end{equation}
Surely as $H^d_G(\omega,\sigma)$ does not contain $\beta$, we have
\begin{align}
 & H^d_G(\emptyset,\emptyset)=\delta_{d,0},\\
 & H^d_G((1),(1))=\delta_{d,0}.
\end{align}\par
When $n\geq 2$, we have
\begin{equation}
\begin{split}
\sum_{d=0}^\infty \beta^d \sum_{\substack{\omega,\sigma\\|\omega|=|\sigma|=n}}H^d_G(\omega,\sigma)=\rho_{n-1}=\prod\limits_{k=1}^{n-2} G(\beta k)
=\prod\limits_{k=1}^{n-2}\prod\limits_{i_k=1}^\infty (1+c_{i_k}k\beta).
\end{split}
\end{equation}\par
For the $N<n$ cases, the equations are underdetermined, we couldn't obtain the value of each $H^d_G(\omega,\sigma)$ with $|\omega|=|\sigma|\geq2$. We have to increase the dimension of the Miwa parameters, such as $N=2$ and $t_k=\frac{1}{k}\sum_{i=1}^{2}\nu_i^{-k},\,s_l=\frac{1}{l}\sum_{j=1}^{2}\mu_j^{-l}$. Then (\ref{tep5-2}) turns into
\begin{equation}\label{tep8}
\begin{split}
  & \sum_{n=0}^\infty\sum_{d=0}^\infty \beta^d \sum_{\substack{\omega,\sigma\\|\omega|=|\sigma|=n}}H^d_G(\omega,\sigma)p_\omega(\bt)p_\sigma(\mathbf{s})\\
= & Term_{-n}\Big\{\frac{r_0(-2)}{(\nu_2-\nu_1)(\mu_2-\mu_1)} \mathop{det}\limits_{i,j=1\cdots 2}\Big(\sum_{l=-2}^\infty \nu_j^{-l-1}\mu_i^{-l-1}\rho_l\Big)\Big\}\\
= & \frac{1}{\rho_{-2}\rho_{-1}}\sum_{\substack{l_1,l_2=-2\\l_1>l_2,l_1+l_2=n-3}}^{\infty} \rho_{l_1}\rho_{l_2}(\nu_1^{-l_1-1}\nu_2^{-l_2-2}+\cdots+\nu_1^{-l_2-2}\nu_2^{-l_1-1})\\
  & \times(\mu_1^{-l_1-1}\mu_2^{-l_2-2}+\cdots+\mu_1^{-l_2-2}\mu_2^{-l_1-1}).
\end{split}
\end{equation}
Let $n=2$, we have four 2 dimensional weighted Hurwitz numbers $H^d_G((2),(2))$, $H^d_G((2),(1,1))$, $H^d_G((1,1),(2))$, $H^d_G((1,1),(1,1))$,
\begin{equation}\label{tep10-1}
\begin{split}
  & \sum_{d=0}^\infty \beta^d \Big(H^d_G((2),(2))(\nu_1^{-2}+\nu_2^{-2}) (\mu_1^{-2}+\mu_2^{-2})+H^d_G((2),(1,1)) (\nu_1^{-2}+\nu_2^{-2})\\
  & \times(\mu_1^{-1}+\mu_2^{-1})(\mu_1^{-1}+\mu_2^{-1})+H^d_G((1,1),(2)) (\nu_1^{-1}+\nu_2^{-1}) (\nu_1^{-1}+\nu_2^{-1})(\mu_1^{-2}+\mu_2^{-2})\\
  & +H^d_G((1,1),(1,1))(\nu_1^{-1}+\nu_2^{-1})(\nu_1^{-1}+\nu_2^{-1}) (\mu_1^{-1}+\mu_2^{-1})(\mu_1^{-1}+\mu_2^{-1})\Big)\\
= & \frac{\rho_{-2}\rho_{1}}{\rho_{-2}\rho_{-1}}(\nu_1^{-2}+\nu_1^{-1}\nu_2^{-1}+\nu_2^{-2}) (\mu_1^{-2} +\mu_1^{-1}\mu_2^{-1}+\mu_2^{-2})+ \frac{\rho_{-1}\rho_{0}}{\rho_{-2}\rho_{-1}} \nu_1^{-1}\nu_2^{-1}\mu_1^{-1}\mu_2^{-1}.
\end{split}
\end{equation}
As $\{\mu_1,\mu_2\}$ and $\{\nu_1,\nu_2\}$ are independent, comparing the coefficients of each side, we have
\begin{align}\label{tep6}
 & H^d_G((2),(2))=H^d_G((1,1),(1,1))=\frac{1}{4}(G_d+(-1)^d G_d),\\
 & H^d_G((2),(1,1))=H^d_G((1,1),(2))=\frac{1}{4}(G_d-(-1)^d G_d).
\end{align}\par
According to the definition of $H^d_G(\omega,\sigma)$ (\ref{def:H_G}), we also can get the values of the 2 dimensional weighted Hurwitz numbers:
\begin{equation}
\begin{split}
H^d_G((1,1),(1,1)) & =W_G(\underbrace{(2),(2),\cdots,(2)}_{\text{the number of }(2)\text{ is }d})H(\underbrace{(2),(2),\cdots,(2)}_{\text{the number of }(2)\text{ is }d},(1,1),(1,1))\\
                   & =\left\{
                   \begin{aligned}
                    & \frac{1}{d!}\sum_{\sigma\in S_d}\sum_{1\leq i_1<\cdots<i_d}c_{i_\sigma(1)}\cdots c_{i_\sigma(d)}\frac{1}{2!} \quad  & d\text{ is even}\\
                    & 0    &d\text{ is odd}
                   \end{aligned}
                   \right.
                   \\
                   & =\frac{1}{4}(G_d+(-1)^d G_d).
\end{split}
\end{equation}\par
In the $n=3$ case, we have
\begin{equation}\label{tep10-2}
\begin{split}
  & \sum_{d=0}^\infty \beta^d H^d_G((3),(3))\nu^{-(3)}\mu^{-(3)}+H^d_G((3),(2,1)) \nu^{-(3)}(\mu^{-(3)}+\mu^{-(2,1)})\\
  & +H^d_G((3),(1,1,1))\nu^{-(3)}(\mu^{-(3)}+3\mu^{-(2,1)}+6\mu^{-(1,1,1)}) +H^d_G((2,1),(3)) (\nu^{-(3)}\\
  & +\nu^{-(2,1)})\mu^{-(3)} +H^d_G((2,1),(2,1))(\nu^{-(3)}+\nu^{-(2,1)}) (\mu^{-(3)} +\mu^{-(2,1)}) \\
  & +H^d_G((2,1),(1,1,1))(\nu^{-(3)}+\nu^{-(2,1)}) (\mu^{-(3)}+3\mu^{-(2,1)}+6\mu^{-(1,1,1)})  \\
  & +H^d_G((1,1,1),(3))(\nu^{-(3)}+3\nu^{-(2,1)}+6\nu^{-(1,1,1)})\mu^{-(3)} +H^d_G((1,1,1),(2,1)) \\
  & \times(\nu^{-(3)}+3\nu^{-(2,1)}+6\nu^{-(1,1,1)})(\mu^{-(3)} +\mu^{-(2,1)})+H^d_G((1,1,1),(1,1,1))  \\
  & \times(\nu^{-(3)}+3\nu^{-(2,1)}+6\nu^{-(1,1,1)}) (\mu^{-(3)}+3\mu^{-(2,1)}+6\mu^{-(1,1,1)})\\
= & \frac{\rho_{0}}{\rho_{-3}}\nu^{-(1,1,1)}\mu^{-(1,1,1)} +\frac{\rho_{1}}{\rho_{-2}}(\nu^{-(2,1)}+2\nu^{-(1,1,1)}) (\mu^{-(2,1)}+2\mu^{-(1,1,1)}) \\
  & \frac{\rho_{2}}{\rho_{-1}}(\nu^{-(3)}+\nu^{-(2,1)}+\nu^{-(1,1,1)}) (\mu^{-(3)}+\mu^{-(2,1)}+\mu^{-(1,1,1)}),
\end{split}
\end{equation}
where $\mu^\lambda=\sum_{\sigma\in \mathbf{S}_n}\mu_1^{\lambda_{\sigma(1)}}\cdots\mu_n^{\lambda_{\sigma(n)}}$ is an abbreviation mentioned in (\ref{eq:abbreviate}). Recall that $\{\mu\}$ and $\{\nu\}$ are independent, comparing the coefficients of each side, we have
\begin{align}\label{tep35}
 & H^d_G((3),(3))=\frac{1}{9}\sum\limits_{k=0}^d 2^k G_k G_{d-k} +\frac{1}{9}\sum\limits_{k=0}^d (-1)^k G_k G_{d-k} +\frac{1}{9}\sum\limits_{k=0}^d (-1)^d 2^k G_k G_{d-k},\\
 & H^d_G((3),(2,1))=H^d_G((2,1),(3))=\frac{1}{6}\sum\limits_{k=0}^d 2^k G_k G_{d-k} -\frac{1}{6}\sum\limits_{k=0}^d (-1)^d 2^k G_k G_{d-k},\notag\\
 & H^d_G((3),(1,1,1))=H^d_G((1,1,1),(3))=\frac{1}{18}\sum\limits_{k=0}^d 2^k G_k G_{d-k} -\frac{1}{9}\sum\limits_{k=0}^d (-1)^k G_k G_{d-k}\notag\\
 &  \qquad\qquad\qquad\qquad+\frac{1}{18}\sum\limits_{k=0}^d (-1)^d 2^k G_k G_{d-k},\notag\\
 & H^d_G((2,1),(2,1))=\frac{1}{4}\sum\limits_{k=0}^d 2^k G_k G_{d-k} +\frac{1}{4}\sum\limits_{k=0}^d (-1)^d 2^k G_k G_{d-k},\notag\\
 & H^d_G((2,1),(1,1,1))=H^d_G((1,1,1),(2,1))=\frac{1}{12}\sum\limits_{k=0}^d 2^k G_k G_{d-k} -\frac{1}{12}\sum\limits_{k=0}^d (-1)^d 2^k G_k G_{d-k},\notag\\
 & H^d_G((1,1,1),(1,1,1))=\frac{1}{36}\sum\limits_{k=0}^d 2^k G_k G_{d-k} +\frac{1}{9}\sum\limits_{k=0}^d (-1)^k G_k G_{d-k} +\frac{1}{36}\sum\limits_{k=0}^d (-1)^d 2^k G_k G_{d-k}.\notag
\end{align}
According to the definition of $H^d_G(\omega,\sigma)$ (\ref{def:H_G}) and $H((3),(3))=\frac{1}{3},\,H((2,1),(2,1),(3))=1$\cite{WJ}, we can calculate $H^2_G((1,1,1),(3))$ to verify (\ref{tep35}):
\begin{equation}
\begin{split}
H^2_G((1,1,1),(3))= & W_G((2,1),(2,1))H((2,1),(2,1),(3))+W_G((3))H((3),(3))\\
                  = & G_2+\frac{1}{3}(G_1G_1-2G_2)\\
                  = & \frac{1}{3}(G_1G_1+G_2),
\end{split}
\end{equation}
and by (\ref{tep35})
\begin{equation}
\begin{split}
H^2_G((1,1,1),(3))= & \frac{1}{18}\sum\limits_{k=0}^2 2^k G_k G_{2-k} -\frac{1}{9}\sum\limits_{k=0}^2 (-1)^k G_k G_{2-k}+\frac{1}{18}\sum\limits_{k=0}^2 (-1)^2 2^k G_k G_{2-k}\\
                  = & \frac{1}{3}(G_1G_1+G_2).
\end{split}
\end{equation}\par
As $n$ increases, the calculation becomes more and more complicated. We give a recursion formula for calculating the higher dimensional weighted Hurwitz number hierarchies.
\begin{LEM}\label{LEM:N_P}
For $P$ a positive integer independent of N
\begin{equation}\label{tep_P}
\begin{split}
 & \mathop{det}\limits_{i,j=1\cdots N}\Big(\sum_{l=-P}^\infty \nu_j^{-l-1}\mu_i^{-l-1}\rho_l\Big)\\
= & \sum_{\substack{l_1,\cdots,l_{N}=-P\\l_1>\cdots>l_N}}^{\infty} \sum_{\sigma\in S_k} \left|\begin{array}{cccc}
\nu_{\sigma(1)}^{-l_1-1}\mu_{1}^{-l_1-1} & \nu_{\sigma(2)}^{-l_2-1}\mu_{1}^{-l_2-1} & \cdots & \nu_{\sigma(N)}^{-l_N-1}\mu_{1}^{-l_N-1}\\
\nu_{\sigma(1)}^{-l_1-1}\mu_{2}^{-l_1-1} & \nu_{\sigma(2)}^{-l_2-1}\mu_{2}^{-l_2-1} & \cdots & \nu_{\sigma(N)}^{-l_N-1}\mu_{2}^{-l_N-1}\\
\vdots & \vdots & \ddots & \vdots\\
\nu_{\sigma(1)}^{-l_1-1}\mu_{N}^{-l_1-1} & \nu_{\sigma(2)}^{-l_2-1}\mu_{N}^{-l_2-1} & \cdots & \nu_{\sigma(N)}^{-l_N-1}\mu_{N}^{-l_N-1}\\
\end{array}\right|sgn(\sigma)\rho_{l_1}\cdots\rho_{l_N}.
\end{split}
\end{equation}
\begin{proof} For each independent $l_i$, the order of summation can be exchanged. We put the summation in the determinant element to the outside of the determinant, and keep the powers of $\mu$ and $\nu$ the same for each column.  Then
\begin{equation}\label{tep_P}
\begin{split}
 & \mathop{det}\limits_{i,j=1\cdots N}\Big(\sum_{l=-P}^\infty \nu_j^{-l-1}\mu_i^{-l-1}\rho_l\Big)\\
= & \sum_{\substack{l_1,\cdots,l_{N}=-P}}^{\infty}\left|\begin{array}{cccc}
\nu_{1}^{-l_1-1}\mu_{1}^{-l_1-1}\rho_{l_1} & \nu_{2}^{-l_2-1}\mu_{1}^{-l_2-1}\rho_{l_2} & \cdots & \nu_{N}^{-l_N-1}\mu_{1}^{-l_N-1}\rho_{l_N}\\
\nu_{1}^{-l_1-1}\mu_{2}^{-l_1-1}\rho_{l_1} & \nu_{2}^{-l_2-1}\mu_{2}^{-l_2-1}\rho_{l_2} & \cdots & \nu_{N}^{-l_N-1}\mu_{2}^{-l_N-1}\rho_{l_N}\\
\vdots & \vdots & \ddots & \vdots\\
\nu_{1}^{-l_1-1}\mu_{N}^{-l_1-1}\rho_{l_1} & \nu_{2}^{-l_2-1}\mu_{N}^{-l_2-1}\rho_{l_2} & \cdots & \nu_{N}^{-l_N-1}\mu_{N}^{-l_N-1}\rho_{l_N}\\
\end{array}\right|.
\end{split}
\end{equation}
For each $\left|\begin{array}{cccc}
\nu_{1}^{-l_1-1}\mu_{1}^{-l_1-1}\rho_{l_1} & \nu_{2}^{-l_2-1}\mu_{1}^{-l_2-1}\rho_{l_2} & \cdots & \nu_{N}^{-l_N-1}\mu_{1}^{-l_N-1}\rho_{l_N}\\
\nu_{1}^{-l_1-1}\mu_{2}^{-l_1-1}\rho_{l_1} & \nu_{2}^{-l_2-1}\mu_{2}^{-l_2-1}\rho_{l_2} & \cdots & \nu_{N}^{-l_N-1}\mu_{2}^{-l_N-1}\rho_{l_N}\\
\vdots & \vdots & \ddots & \vdots\\
\nu_{1}^{-l_1-1}\mu_{N}^{-l_1-1}\rho_{l_1} & \nu_{2}^{-l_2-1}\mu_{N}^{-l_2-1}\rho_{l_2} & \cdots & \nu_{N}^{-l_N-1}\mu_{N}^{-l_N-1}\rho_{l_N}\\
\end{array}\right|$, if there exists a determinant for which $l_i=l_j,\,i\neq j$, then the $i$-th column of the determinant is proportional to the $j$-th column, which means the  determinant vanishes. While for the remaining determinants whose $l_1, \cdots, l_N$ are different from each other, we do elementary column transformations on them to
make the power of $\mu$ and $\nu$ to be arranged from small to large, and we then have
\begin{align}\label{tep_P}
& \mathop{det}\limits_{i,j=1\cdots N}\Big(\sum_{l=-P}^\infty \nu_j^{-l-1}\mu_i^{-l-1}\rho_l\Big)\\
= & \sum_{\substack{l_1,\cdots,l_{N}=-P\\l_1>\cdots>l_N}}^{\infty}\sum_{\sigma\in S_k}sgn(\sigma)\left|\begin{array}{cccc}
\nu_{\sigma(1)}^{-l_1-1}\mu_{1}^{-l_1-1}\rho_{l_1} & \nu_{\sigma(2)}^{-l_2-1}\mu_{1}^{-l_2-1}\rho_{l_2} & \cdots & \nu_{\sigma(N)}^{-l_N-1}\mu_{1}^{-l_N-1}\rho_{l_N}\\
\nu_{\sigma(1)}^{-l_1-1}\mu_{2}^{-l_1-1}\rho_{l_1} & \nu_{\sigma(2)}^{-l_2-1}\mu_{2}^{-l_2-1}\rho_{l_2} & \cdots & \nu_{\sigma(N)}^{-l_N-1}\mu_{2}^{-l_N-1}\rho_{l_N}\\
\vdots & \vdots & \ddots & \vdots\\
\nu_{\sigma(1)}^{-l_1-1}\mu_{N}^{-l_1-1}\rho_{l_1} & \nu_{\sigma(2)}^{-l_2-1}\mu_{N}^{-l_2-1}\rho_{l_2} & \cdots & \nu_{\sigma(N)}^{-l_N-1}\mu_{N}^{-l_N-1}\rho_{l_N}\\
\end{array}\right|\notag\\
= & \sum_{\substack{l_1,\cdots,l_{N}=-P\\l_1>\cdots>l_N}}^{\infty}\sum_{\sigma\in S_k}sgn(\sigma) \left|\begin{array}{cccc}
\nu_{\sigma(1)}^{-l_1-1}\mu_{1}^{-l_1-1} & \nu_{\sigma(2)}^{-l_2-1}\mu_{1}^{-l_2-1} & \cdots & \nu_{\sigma(N)}^{-l_N-1}\mu_{1}^{-l_N-1}\\
\nu_{\sigma(1)}^{-l_1-1}\mu_{2}^{-l_1-1} & \nu_{\sigma(2)}^{-l_2-1}\mu_{2}^{-l_2-1} & \cdots & \nu_{\sigma(N)}^{-l_N-1}\mu_{2}^{-l_N-1}\\
\vdots & \vdots & \ddots & \vdots\\
\nu_{\sigma(1)}^{-l_1-1}\mu_{N}^{-l_1-1} & \nu_{\sigma(2)}^{-l_2-1}\mu_{N}^{-l_2-1} & \cdots & \nu_{\sigma(N)}^{-l_N-1}\mu_{N}^{-l_N-1}\\
\end{array}\right|\rho_{l_1}\cdots\rho_{l_N}.\notag
\end{align}
\end{proof}
\end{LEM}
\begin{LEM}\label{LEM:recursion}
For any positive integer $k\geq2$, Miwa parameters $\hat{t}_m=\frac{1}{m}\sum_{i=1}^{N+1}\nu_i^{-m},\, \hat{s}_r=\frac{1}{r}\sum_{j=1}^{N+1}\mu_j^{-r}$ and $1\leq N\leq k$, we have
\begin{align}\label{tep33}
  & Term_{-k-1}\Big\{\frac{r_0(-(N+1))}{\Delta(\nu_1\cdots\nu_{N+1}) \Delta(\mu_1\cdots\mu_{N+1})}\mathop{det}\limits_{i,j=1\cdots N+1}\Big(\sum_{l=-N-1}^\infty \nu_j^{-l-1}\mu_i^{-l-1}\rho_l\Big)\Big\}\\
= & \sum_{{\alpha_1}=1}^{N+1} \frac{\mu_{{\alpha_1}}^{N}} {\prod\limits_{\substack{i=1\\i\neq{\alpha_1}}}^{N+1}(\mu_{\alpha_1}-\mu_i)}  \sum_{{\beta_1}=1}^{N+1}\frac{\nu_{{\beta_1}}^{N}} {\prod\limits_{\substack{j=1\\j\neq{\beta_1}}} ^{N+1}(\nu_{\beta_1}-\nu_j)} Term_{-k-1}\Big\{\mathop{det}\limits_{\substack{i,j=1\cdots N+1\\i\neq {\alpha_1} ,\, j\neq {\beta_1}}}\Big(\sum_{l=-N}^\infty \nu_j^{-l-1}\mu_i^{-l-1}\rho_l\Big)\notag\\
  & \times\frac{r_0(-N)}{\Delta(\{\mu_1, \cdots,\hat{\mu}_{\alpha_1},\cdots,\mu_{N+1}\}) \Delta(\{\nu_1,\cdots,\hat{\nu}_{\beta_1},\cdots,\nu_{N+1}\})} \Big\}\notag\\
  & + \sum_{l_{N+1}=-N}^{k-N}\frac{\rho_{l_{N+1}}}{\rho_{-N-1}} \sum_{\substack{\omega,\sigma\\|\omega|=|\sigma|=k-l_{N+1}-N}} \bar{p}_{\sigma}(\hat{\bt},1,l_{N+1})\bar{p}_{\omega}(\hat{\bs},1,l_{N+1}) \sum_{d=0}^\infty {\beta}^d r_0(-N)\notag\\
  & \times Term_{(\rho_j,\,j\geq l_{N+1}+1)}\Big\{\frac{H^d_G(\omega,\sigma)} {r_0(-N)}\Big\},\notag
\end{align}
where $Term_{(\rho_j,\,j\geq l_{N+1}+1)}\{f(\{\rho_j\})\}$ denotes the terms $\rho_{l_1},\,\cdots,\,\rho_{l_N}$, whose subscripts $\{l_1,\,\cdots,\,l_{N}\}$ are greater than or equal to $l_{N+1}+1$, in the function $f(\{\rho_j\})$ which is a polynomial with respect to $\{\rho_j\}_{j\in\mathbb{Z}}$ and $\bar{p}_{\sigma}(\hat{\bt},1,l_{N+1})=\sum_{{\alpha_1}=1}^{N+1} \frac{\mu_{{\alpha_1}}^{-l_{N+1}-1}} {\prod\limits_{\substack{i=1\\i\neq{\alpha_1}}}^{N+1}(\mu_{\alpha_1}-\mu_i)} p_\sigma(\{\mu_1,\cdots,\hat{\mu}_{\alpha_1},\cdots,\mu_{N+1}\})$ and $p_\sigma=p_{\sigma_1}\cdots p_{\sigma_{k-l_{N+1}-N}}$, $ p_r(x_1,\cdots,x_n)=\sum_{i=1}^n x_{i}^r$ is the power sum\cite{Mac}.
\begin{proof}
Let us focus on the determinant of the left hand side of (\ref{tep33}).
\begin{align}\label{tep_k+1}
 & \mathop{det}\limits_{i,j=1\cdots N+1}\Big(\sum_{l=-N-1}^\infty \nu_j^{-l-1}\mu_i^{-l-1}\rho_l\Big)\\
= & \sum_{\substack{l_1,\cdots,l_{N+1}=-N-1\\l_1>\cdots>l_{N+1}}}^{\infty} \sum_{\sigma\in S_{N+1}}sgn(\sigma)\left|\begin{array}{cccc}
\nu_{\sigma(1)}^{-l_1-1}\mu_{1}^{-l_1-1}\rho_{l_1} & \nu_{\sigma(2)}^{-l_2-1}\mu_{1}^{-l_2-1}\rho_{l_2} & \cdots & \nu_{\sigma(N+1)}^{-l_{N+1}-1}\mu_{1}^{-l_{N+1}-1}\rho_{l_{N+1}}\\
\nu_{\sigma(1)}^{-l_1-1}\mu_{2}^{-l_1-1}\rho_{l_1} & \nu_{\sigma(2)}^{-l_2-1}\mu_{2}^{-l_2-1}\rho_{l_2} & \cdots & \nu_{\sigma(N+1)}^{-l_{N+1}-1}\mu_{2}^{-l_{N+1}-1}\rho_{l_{N+1}}\\
\vdots & \vdots & \ddots & \vdots\\
\nu_{\sigma(1)}^{-l_1-1}\mu_{N+1}^{-l_1-1}\rho_{l_1} & \nu_{\sigma(2)}^{-l_2-1}\mu_{N+1}^{-l_2-1}\rho_{l_2} & \cdots & \nu_{\sigma(N+1)}^{-l_{N+1}-1}\mu_{N+1}^{-l_{N+1}-1}\rho_{l_{N+1}}\\
\end{array}\right|\notag\\
= & \sum_{l_{N+1}=-N-1}^{\infty}\rho_{l_{N+1}} \sum_{\substack{l_1,\cdots,l_{N+1}=l_{N+1}+1\\l_1>\cdots>l_{N}}}^{\infty} \rho_{l_1}\cdots\rho_{l_N}\sum_{\sigma\in S_{N+1}}sgn(\tilde{\sigma})(-1)^{\sigma(N+1)+N+1}\sum_{{\alpha_1}=1}^{N+1} \nu_{\sigma(N+1)}^{-l_{N+1}-1}\mu_{{\alpha_1}}^{-l_{N+1}-1}\notag\\
  & (-1)^{{\alpha_1}+N+1} \left|\begin{array}{cccc}
\nu_{\sigma(1)}^{-l_1-1}\mu_{1}^{-l_1-1}\rho_{l_1} & \nu_{\sigma(2)}^{-l_2-1}\mu_{1}^{-l_2-1}\rho_{l_2} & \cdots & \nu_{\sigma(N+1)}^{-l_{N+1}-1}\mu_{1}^{-l_{N+1}-1}\rho_{l_{N+1}}\\
\vdots & \vdots & \vdots & \vdots\\
\nu_{\sigma(1)}^{-l_1-1}\mu_{{\alpha_1}-1}^{-l_1-1}\rho_{l_1} & \nu_{\sigma(2)}^{-l_2-1}\mu_{{\alpha_1}-1}^{-l_2-1}\rho_{l_2} & \cdots & \nu_{\sigma(N+1)}^{-l_{N+1}-1}\mu_{{\alpha_1}-1}^{-l_{N+1}-1}\rho_{l_{N+1}}\\
\nu_{\sigma(1)}^{-l_1-1}\mu_{{\alpha_1}+1}^{-l_1-1}\rho_{l_1} & \nu_{\sigma(2)}^{-l_2-1}\mu_{{\alpha_1}+1}^{-l_2-1}\rho_{l_2} & \cdots & \nu_{\sigma(N+1)}^{-l_{N+1}-1}\mu_{{\alpha_1}+1}^{-l_{N+1}-1}\rho_{l_{N+1}}\\
\vdots & \vdots & \ddots & \vdots\\
\nu_{\sigma(1)}^{-l_1-1}\mu_{N+1}^{-l_1-1}\rho_{l_1} & \nu_{\sigma(2)}^{-l_2-1}\mu_{N+1}^{-l_2-1}\rho_{l_2} & \cdots & \nu_{\sigma(N+1)}^{-l_{N+1}-1}\mu_{N+1}^{-l_{N+1}-1}\rho_{l_{N+1}}\\
\end{array}\right|,\notag
\end{align}
where $\tilde{\sigma}$ is equal to removing $\sigma(N+1)$ from $\sigma$ and subtracting 1 from the items greater than $\sigma(N+1)$, which means
\begin{equation}
\tilde{\sigma}(i)=\left\{
                   \begin{aligned}
                    & \sigma(i) \quad  & \sigma(i)<\sigma(N+1),\\
                    & \sigma(i)-1 \quad  & \sigma(i)>\sigma(N+1).\\
                   \end{aligned}
                   \right.
\end{equation}\
So $\tilde{\sigma}$ is a permutation in $S_k$ and $sgn(\tilde{\sigma})=(-1)^{\sigma(N+1)+N+1}$ $ sgn(\sigma)$. By Lemma [\ref{LEM:N_P}], we have
\begin{align}\label{tep28}
 & \frac{r_0(-N-1)}{\Delta(\{\nu_j\}_{j=1}^{N+1}) \Delta(\{\mu_i\}_{i=1}^{N+1})}\mathop{det}\limits_{i,j=1\cdots N+1}\Big(\sum_{l=-N-1}^\infty \nu_j^{-l-1}\mu_i^{-l-1}\rho_l\Big)\\
= & \frac{r_0(-N-1)}{\Delta(\{\nu_j\}_{j=1}^{N+1})\Delta(\{\mu_i\}_{i=1}^{N+1})} \sum_{l_{N+1}=-N-1}^\infty\rho_{l_{N+1}} \sum_{{\alpha_1}=1}^{N+1} \sum_{{\beta_1}=1}^{N+1}(-1)^{{\beta_1}+N+1}\nu_{{\beta_1}}^{-l_{N+1}-1} \mu_{{\alpha_1}}^{l_{N+1}-1}\notag\\
 & \times(-1)^{{\alpha_1}+N+1}\mathop{det}\limits_{\substack{i,j=1\cdots N+1\\i\neq {\alpha_1} ,\, j\neq {\beta_1}}}\Big(\sum_{l=l_{N+1}+1}^\infty \nu_j^{-l-1}\mu_i^{-l-1}\rho_l\Big)\notag\\
= & \sum_{{\alpha_1}=1}^{N+1}\sum_{{\beta_1}=1}^{N+1}\frac{r_0(-N)} {\Delta(\{\nu_1,\cdots,\hat{\nu}_{\beta_1},\cdots,\nu_{N+1}\}) \Delta(\{\mu_1,\cdots,\hat{\mu}_{\alpha_1},\cdots,\mu_{N+1}\})}\notag\\
  & \times\sum_{l_{N+1}=-N-1}^\infty\frac{\rho_{l_{N+1}}}{\rho_{-N-1}} \frac{\nu_{{\beta_1}}^{-l_{N+1}-1}}{\prod\limits_{\substack{j=1\\j\neq{\beta_1}}} ^{N+1}(\nu_{\beta_1}-\nu_j)} \frac{\mu_{{\alpha_1}}^{-l_{N+1}-1}}{\prod\limits_{\substack{i=1\\i\neq{\alpha_1}}}^{N+1}(\mu_{\alpha_1}-\mu_i)} \mathop{det}\limits_{\substack{i,j=1\cdots N+1\\i\neq {\alpha_1} ,\, j\neq {\beta_1}}}\Big(\sum_{l=l_{N+1}+1}^\infty \nu_j^{-l-1}\mu_i^{-l-1}\rho_l\Big).\notag
\end{align}\par
To get the value of $H^d_G(\omega,\sigma),\, |\omega|=|\sigma|=k+1$, we only need the term of power $-k-1$ which means the powers of $\{\mu_i\}_{i=1}^N$ and $\{\nu_j\}_{j=1}^N$ are $-k-1$. According to the Taylor expansion, when $|\mu_i|>|\mu_j|$, we have
\begin{equation}
\begin{split}
  & \frac{1}{\mu_i-\mu_j}=\frac{1}{\mu_i}\times\frac{1}{1-\frac{\mu_j}{\mu_i}}
=\frac{1}{\mu_i}+\frac{\mu_j}{\mu_i^2}+\frac{\mu_j^2}{\mu_i^3}+\cdots,
\end{split}
\end{equation}
and when $|\mu_i|<|\mu_j|$, we have
\begin{equation}
\begin{split}
  & \frac{1}{\mu_i-\mu_j}=-\frac{1}{\mu_j}\times\frac{1}{1-\frac{\mu_i}{\mu_j}}
=-\frac{1}{\mu_j}-\frac{\mu_i}{\mu_j^2}-\frac{\mu_i^2}{\mu_j^3}+\cdots.
\end{split}
\end{equation}
Assuming $|\mu_1|>|\mu_2|>\cdots>|\mu_{k+1}|$, the sum of the powers of $\nu$ and $\mu$ in $\frac{\nu_{{\beta_1}}^{-l_{N+1}-1}}{\prod\limits_{\substack{j=1\\j\neq{\beta_1}}} ^{N+1}(\nu_{\beta_1}-\nu_j)}$ and in $\frac{\mu_{{\alpha_1}}^{-l_{N+1}-1}}{\prod\limits_{\substack{i=1\\i\neq{\alpha_1}}}^{N+1}(\mu_{\alpha_1}-\mu_i)}$ are both $-l_{N+1}-N-1$. Therefore, $l_{N+1}$ not only satisfies $l_{N+1}\geq-N-1$ but also $l_{N+1}\leq k-N$ after acting by $Term_{-k-1}\{\cdots\}$. If $l_{k+1}>k-N$, then the sum of the powers of the monomials in (\ref{tep28}) is $<(-k-1)(N+1)$ which vanishes under $Term_{-k-1}\{\cdots\}$.
\begin{align}\label{tep29}
  & Term_{-k-1}\Big\{\frac{r_0(-(N+1))}{\Delta(\nu_1\cdots\nu_{N+1}) \Delta(\mu_1\cdots\mu_{N+1})}\mathop{det}\limits_{i,j=1\cdots N+1}\Big(\sum_{l=-N-1}^\infty \nu_j^{-l-1}\mu_i^{-l-1}\rho_l\Big)\Big\}\\
= & Term_{-k-1}\Big\{\sum_{l_{N+1}=-N-1}^{k-N}\frac{\rho_{l_{N+1}}} {\rho_{-N-1}} \sum_{{\alpha_1}=1}^{N+1}\sum_{{\beta_1}=1}^{N+1} \frac{\nu_{{\beta_1}}^{-l_{N+1}-1}} {\prod\limits_{\substack{j=1\\ j\neq{\beta_1}}} ^{N+1}(\nu_{\beta_1}-\nu_j)} \frac{\mu_{{\alpha_1}} ^{-l_{N+1}-1}}{\prod\limits_{\substack{i=1\\i\neq{\alpha_1}}} ^{N+1}(\mu_{\alpha_1}-\mu_i)}\notag\\
  & \times \frac{r_0(-N)}{\Delta(\{\nu_1,\cdots,\hat{\nu}_{\beta_1}, \cdots,\nu_{N+1}\}) \Delta(\{\mu_1, \cdots,\hat{\mu}_{\alpha_1},\cdots,\mu_{N+1}\})}\notag\\
  & \times\mathop{det}\limits_{\substack{i,j=1\cdots N+1\\i\neq {\alpha_1} ,\, j\neq {\beta_1}}}\Big(\sum_{l=l_{N+1}+1}^\infty \nu_j^{-l-1}\mu_i^{-l-1}\rho_l\Big)\Big\}\notag\\
= & \sum_{l_{N+1}=-N-1}^{k-N}\frac{\rho_{l_{N+1}}}{\rho_{-N-1}} \sum_{{\alpha_1}=1}^{N+1}\sum_{{\beta_1}=1}^{N+1} \frac{\nu_{{\beta_1}}^{-l_{N+1}-1}}{\prod\limits_{\substack{j=1\\j\neq{\beta_1}}} ^{N+1}(\nu_{\beta_1}-\nu_j)} \frac{\mu_{{\alpha_1}}^{-l_{N+1}-1}} {\prod\limits_{\substack{i=1\\i\neq{\alpha_1}}}^{N+1} (\mu_{\alpha_1}-\mu_i)}\notag\\
  & \times Term_{-k+l_{N+1}+N}\Big\{\frac{r_0(-N)} {\Delta(\{\nu_1,\cdots,\hat{\nu}_{\beta_1},\cdots,\nu_{N+1}\}) \Delta(\{\mu_1, \cdots,\hat{\mu}_{\alpha_1},\cdots,\mu_{N+1}\})}\notag\\
  & \times\mathop{det}\limits_{\substack{i,j=1\cdots N+1\\i\neq {\alpha_1} ,\, j\neq {\beta_1}}}\Big(\sum_{l=l_{N+1}+1}^\infty \nu_j^{-l-1}\mu_i^{-l-1}\rho_l\Big)\Big\}.\notag
\end{align}\par
Let $t_m=\frac{1}{m}\sum\limits_{i=1}^{N}\nu_i^{-m},\,s_r =\frac{1}{r}\sum\limits_{j=1}^{N}\mu_j^{-r}$ and $n=k-l_{N+1}-N$, then (\ref{tep5-2}) turn into
\begin{equation}\label{tep30}
\begin{split}
  & \sum_{\substack{\omega,\sigma\\|\omega|=|\sigma|=k-l_{N+1}-N}}\sum_{d=0}^\infty {\beta}^d \frac{H^d_G(\omega,\sigma)} {r_0(-N)} p_\omega(\bt)p_\sigma(\mathbf{s})\\
= & Term_{-k+l_{N+1}+N}\Big\{\frac{1} {\Delta(\nu)\Delta(\mu)}\mathop{det}\limits_{i,j=1\cdots N}\Big(\sum_{l=-N}^\infty \nu_j^{-l-1}\mu_i^{-l-1}\rho_l\Big)\Big\}.
\end{split}
\end{equation}
For the subscript of $\rho_l$ being a negative power of $\nu$ or $\mu -1$ in $\nu_j^{-l-1}\mu_i^{-l-1}\rho_l$, take the item whose subscripts $\{l_1,\,\cdots,\,l_{N}\}$ of $\rho_{l_1}\cdots\rho_{l_N}$ are greater than or equal to $l_{N+1}+1$ on both sides of (\ref{tep30}), and we have
\begin{equation}\label{tep32}
\begin{split}
  &  \sum_{\substack{\omega,\sigma\\|\omega|=|\sigma|=k-l_{N+1}-N}} \sum_{d=0}^\infty {\beta}^d r_0(-N) Term_{(\rho_j,\,j\geq l_{N+1}+1)}\Big\{\frac{H^d_G(\omega,\sigma)} {r_0(-N)}\Big\} p_\omega(\bt)p_\sigma(\mathbf{s})\\
&~~~~= Term_{-k+l_{N+1}+N}\Big\{\frac{r_0(-N)} {\Delta(\nu)\Delta(\mu)}\mathop{det}\limits_{i,j=1\cdots N}\Big(\sum_{l=l_{N+1}+1}^\infty \nu_j^{-l-1}\mu_i^{-l-1}\rho_l\Big)\Big\}.
\end{split}
\end{equation}
Substituting (\ref{tep32}) into (\ref{tep29}), we have
\begin{align}
  & Term_{-k-1}\Big\{\frac{r_0(-(N+1))}{\Delta(\nu_1\cdots\nu_{N+1}) \Delta(\mu_1\cdots\mu_{N+1})}\mathop{det}\limits_{i,j=1\cdots N+1}\Big(\sum_{l=-N-1}^\infty \nu_j^{-l-1}\mu_i^{-l-1}\rho_l\Big)\Big\}\\
= & \sum_{{\alpha_1}=1}^{N+1} \frac{\mu_{{\alpha_1}}^{N}} {\prod\limits_{\substack{i=1\\i\neq{\alpha_1}}}^{N+1}(\mu_{\alpha_1}-\mu_i)}  \sum_{{\beta_1}=1}^{N+1}\frac{\nu_{{\beta_1}}^{N}} {\prod\limits_{\substack{j=1\\j\neq{\beta_1}}} ^{N+1}(\nu_{\beta_1}-\nu_j)} Term_{-k-1}\Big\{\mathop{det}\limits_{\substack{i,j=1\cdots N+1\\i\neq {\alpha_1} ,\, j\neq {\beta_1}}}\Big(\sum_{l=-N}^\infty \nu_j^{-l-1}\mu_i^{-l-1}\rho_l\Big)\notag\\
  & \times\frac{r_0(-N)}{\Delta(\{\mu_1, \cdots,\hat{\mu}_{\alpha_1},\cdots,\mu_{N+1}\}) \Delta(\{\nu_1,\cdots,\hat{\nu}_{\beta_1},\cdots,\nu_{N+1}\})}\Big\}\notag\\
  & + \sum_{l_{N+1}=-N}^{k-N}\frac{\rho_{l_{N+1}}}{\rho_{-N-1}} \sum_{\substack{\omega,\sigma\\|\omega|=|\sigma|=k-l_{N+1}-N}} \sum_{{\alpha_1}=1}^{N+1}\frac{\mu_{{\alpha_1}}^{-l_{N+1}-1}} {\prod\limits_{\substack{i=1\\i\neq{\alpha_1}}}^{N+1}(\mu_{\alpha_1}-\mu_i)} p_\sigma(\{\mu_1,\cdots,\hat{\mu}_{\alpha_1},\cdots,\mu_{k+1}\})\notag\\
  & \times\sum_{{\beta_1}=1}^{N+1}\frac{\nu_{{\beta_1}}^{-l_{N+1}-1}} {\prod\limits_{\substack{j=1\\j\neq{\beta_1}}} ^{N+1}(\nu_{\beta_1}-\nu_j)} p_\omega(\{\nu_1,\cdots,\hat{\nu}_{\beta_1},\cdots,\nu_{k+1}\}) \sum_{d=0}^\infty {\beta}^d r_0(-N) Term_{(\rho_j,\,j\geq l_{N+1}+1)}\Big\{\frac{H^d_G(\omega,\sigma)} {r_0(-N)}\Big\}\notag\\
= & \sum_{{\alpha_1}=1}^{N+1} \frac{\mu_{{\alpha_1}}^{N}} {\prod\limits_{\substack{i=1\\i\neq{\alpha_1}}}^{N+1}(\mu_{\alpha_1}-\mu_i)}  \sum_{{\beta_1}=1}^{N+1}\frac{\nu_{{\beta_1}}^{N}} {\prod\limits_{\substack{j=1\\j\neq{\beta_1}}} ^{N+1}(\nu_{\beta_1}-\nu_j)} Term_{-k-1}\Big\{\mathop{det}\limits_{\substack{i,j=1\cdots N+1\\i\neq {\alpha_1} ,\, j\neq {\beta_1}}}\Big(\sum_{l=-N}^\infty \nu_j^{-l-1}\mu_i^{-l-1}\rho_l\Big)\notag\\
  & \times\frac{r_0(-N)}{\Delta(\{\mu_1, \cdots,\hat{\mu}_{\alpha_1},\cdots,\mu_{N+1}\}) \Delta(\{\nu_1,\cdots,\hat{\nu}_{\beta_1},\cdots,\nu_{N+1}\})}\Big\}+ \sum_{l_{N+1}=-N}^{k-N}\frac{\rho_{l_{N+1}}}{\rho_{-N-1}}\notag\\
  &  \times\sum_{\substack{\omega,\sigma\\|\omega|=|\sigma|=k-l_{N+1}-N}} \bar{p}_{\sigma}(\hat{\bt},1,l_{N+1})\bar{p}_{\omega}(\hat{\bs},1,l_{N+1}) \sum_{d=0}^\infty {\beta}^d r_0(-N) Term_{(\rho_j,\,j\geq l_{N+1}+1)}\Big\{\frac{H^d_G(\omega,\sigma)} {r_0(-N)}\Big\},\notag
\end{align}
where $\bar{p}_{\sigma}(\hat{\bt},1,l_{N+1})=\sum_{{\alpha_1}=1}^{N+1} \frac{\mu_{{\alpha_1}}^{-l_{N+1}-1}} {\prod\limits_{\substack{i=1\\i\neq{\alpha_1}}}^{N+1}(\mu_{\alpha_1}-\mu_i)} p_\sigma(\{\mu_1,\cdots,\hat{\mu}_{\alpha_1},\cdots,\mu_{N+1}\})$.
\end{proof}
\end{LEM}
\begin{THM}\label{THM:recursion}For any positive integer $k\geq2$ and Miwa parameters $\hat{t}_m=\frac{1}{m}\sum_{i=1}^{N+1}\nu_i^{-m},\, \hat{s}_r=\frac{1}{r}\sum_{j=1}^{N+1}\mu_j^{-r}$, we have
\begin{align}\label{tep34}
  & \sum_{d=0}^\infty \beta^d \sum_{\substack{\omega,\sigma\\|\omega|=|\sigma|=k+1}}H^d_G(\omega,\sigma) p_\omega(\bt)p_\sigma(\mathbf{s})\\
= & \frac{\rho_{k}}{\rho_{-1}}\frac{\sum\limits_{\zeta\in S_{k+1}}sgn(\zeta)\mu_{\zeta(1)}^k \cdots\mu_{\zeta(k)}\mu_{\zeta(k+1)}^{-k-1}} {\Delta(\mu)}\frac{\sum\limits_{\tilde{\zeta}\in S_{k+1}}sgn(\tilde{\zeta})\nu_{\tilde{\zeta}(1)}^k \cdots\nu_{\tilde{\zeta}(k)}\nu_{\tilde{\zeta}(k+1)}^{-k-1}}{\Delta(\nu)}\notag \\
  & + \sum_{l_{2}=-1}^{k-1}\frac{\rho_{l_{2}}}{\rho_{-2}} \sum_{\substack{\omega,\sigma\\|\omega|=|\sigma|=k-l_{2}-1}} \bar{p}_{\sigma}(\bt,k-1,l_2) \bar{p}_{\omega}(\bs,k-1,l_2)\notag\\
  & \times \sum_{d=0}^\infty {\beta}^d Term_{(\rho_j,\,j\geq l_{2}+1)}\{H^d_G(\omega,\sigma)\}+ \cdots\notag\\
  & + \sum_{l_{k+1}=-k}^{0}\frac{\rho_{l_{k+1}}}{\rho_{-k-1}} \sum_{\substack{\omega,\sigma\\|\omega|=|\sigma|=-l_{k+1}}} \bar{p}_{\sigma}(\bt,1,l_{k+1})\bar{p}_{\omega}(\bs,1,l_{k+1})\notag\\
  & \times \sum_{d=0}^\infty {\beta}^d r_0(-k) Term_{(\rho_j,\,j\geq l_{k+1}+1)}\Big\{\frac{H^d_G(\omega,\sigma)} {r_0(-k)}\Big\},\notag
\end{align}
where $\bar{p}_{\sigma}(\bt,m,l)=\sum_{\substack{\alpha_1,\cdots,\alpha_m=1\\ \alpha_i\neq\alpha_j,i\neq j}}^{k+1} \frac{\mu_{\alpha_1}^{k}\mu_{\alpha_2}^{k-1} \cdots \mu_{\alpha_{m-1}}^{k+2-m}\mu_{\alpha_m}^{-l-1}} {\prod\limits_{\substack{i=1\\i\neq\alpha_1}}^{k+1}(\mu_{\alpha_1}-\mu_i) \prod\limits_{\substack{i=1\\i\neq\alpha_1,\alpha_2}}^{k+1}(\mu_{\alpha_2}-\mu_i) \cdots \prod\limits_{\substack{i=1\\i\neq\alpha_1,\cdots,\alpha_m}}^{k+1} (\mu_{\alpha_2}-\mu_i)}$\\
$~~~~~~~~~~~~~~~~~~~~~~~~~\times p_\sigma(\{\mu_1,\cdots,\hat{\mu}_{\alpha_1},\cdots, \hat{\mu}_{\alpha_m},\cdots,\mu_{k+1}\})$.
\begin{proof}
From (\ref{tep9}), we have
\begin{equation}\nonumber
\sum_{d=0}^\infty \beta^d \sum_{\substack{\omega,\sigma\\|\omega|=|\sigma|=n}} H^d_G(\omega,\sigma)=\rho_{n-1}.
\end{equation}
And if we want to obtain the value of $H^d_G(\omega,\sigma),\,|\omega|=|\sigma|=k+1$, we need to make the $n$ in (\ref{tep5-1}) equal to $k+1$ and make $N$
greater than or equal to $k+1$. For the convenience of calculation, we use $N=n=k+1$,
\begin{equation}\nonumber
\begin{split}
  & \sum_{d=0}^\infty \beta^d \sum_{\substack{\omega,\sigma\\|\omega|=|\sigma|=k+1}}H^d_G(\omega,\sigma) p_\omega(\bt)p_\sigma(\mathbf{s}) =Term_{-k-1}\Big\{\frac{r_0(-k-1)}{\Delta(\nu)\Delta(\mu)}\mathop{det}\limits_{i,j=1\cdots k+1}\Big(\sum_{l=-k-1}^\infty \nu_j^{-l-1}\mu_i^{-l-1}\rho_l\Big)\Big\},
\end{split}
\end{equation}
where $t_m=\frac{1}{m}\sum_{i=1}^{k+1}\nu_i^{-m}, \,s_r=\frac{1}{r}\sum_{j=1}^{k+1}\mu_j^{-r}$. Using [Lemma \ref{LEM:recursion}], we have
\begin{align}
  & \,\,\quad\sum_{d=0}^\infty \beta^d \sum_{\substack{\omega,\sigma\\|\omega|=|\sigma|=k+1}}H^d_G(\omega,\sigma) p_\omega(\bt)p_\sigma(\mathbf{s})\\
= & \sum_{{\alpha_1}=1}^{k+1} \frac{\mu_{{\alpha_1}}^{k}} {\prod\limits_{\substack{i=1\\i\neq{\alpha_1}}}^{k+1}(\mu_{\alpha_1}-\mu_i)}  \sum_{{\beta_1}=1}^{k+1}\frac{\nu_{{\beta_1}}^{k}} {\prod\limits_{\substack{j=1\\j\neq{\beta_1}}} ^{k+1}(\nu_{\beta_1}-\nu_j)} Term_{-k-1}\Big\{\mathop{det}\limits_{\substack{i,j=1\cdots k+1\\i\neq {\alpha_1} ,\, j\neq {\beta_1}}}\Big(\sum_{l=-k}^\infty \nu_j^{-l-1}\mu_i^{-l-1}\rho_l\Big)\notag\\
  & \times\frac{r_0(-k)}{\Delta(\{\mu_1, \cdots,\hat{\mu}_{\alpha_1},\cdots,\mu_{k+1}\}) \Delta(\{\nu_1,\cdots,\hat{\nu}_{\beta_1},\cdots,\nu_{k+1}\})} \Big\}\notag\\
  & + \sum_{l_{k+1}=-k}^{0}\frac{\rho_{l_{k+1}}}{\rho_{-k-1}} \sum_{\substack{\omega,\sigma\\|\omega|=|\sigma|=-l_{k+1}}} \bar{p}_{\sigma}(\bt,\mu_{\alpha_1})\bar{p}_{\omega}(\bs,\nu_{\beta_1}) \sum_{d=0}^\infty {\beta}^d r_0(-k) Term_{(\rho_j,\,j\geq l_{k+1}+1)}\Big\{\frac{H^d_G(\omega,\sigma)} {r_0(-k)}\Big\}\notag\\
= & \sum_{\substack{\alpha_1,\alpha_2=1\\ \alpha_2\neq\alpha_1}}^{k+1} \frac{\mu_{{\alpha_1}}^{k}\mu_{{\alpha_2}}^{k-1}} {\prod\limits_{\substack{i=1\\i\neq{\alpha_1}}}^{k+1}(\mu_{\alpha_1}-\mu_i) \prod\limits_{\substack{i=1\\i\neq{\alpha_1,\alpha_2}}}^{k+1} (\mu_{\alpha_2}-\mu_i)} \sum_{\substack{{\beta_1,\beta_2}=1\\ \beta_1\neq\beta_2}}^{k+1}\frac{\nu_{{\beta_1}}^{k}\nu_{\beta_2}^{k-1}} {\prod\limits_{\substack{j=1\\j\neq{\beta_1}}} ^{k+1}(\nu_{\beta_1}-\nu_j) \prod\limits_{\substack{j=1\\j\neq{\beta_1,\beta_2}}} ^{k+1}(\nu_{\beta_2}-\nu_j)}\notag\\
  & \times Term_{-k-1}\Big\{\mathop{det}\limits_{\substack{i,j=1\cdots k+1\\i\neq {\alpha_1,\alpha_2} ,\, j\neq {\beta_1,\beta_2}}}\Big(\sum_{l=-k+1}^\infty \nu_j^{-l-1}\mu_i^{-l-1}\rho_l\Big) \notag\\
  & \times\frac{r_0(-k+1)}{\Delta(\{\mu_1, \cdots,\hat{\mu}_{\alpha_1},\cdots,\hat{\mu}_{\alpha_2},\cdots,\mu_{k+1}\}) \Delta(\{\nu_1,\cdots,\hat{\nu}_{\beta_1},\cdots,\hat{\nu}_{\beta_2},\cdots, \nu_{k+1}\})} \Big\}\notag\\
  & + \sum_{l_{k}=-k+1}^{1}\frac{\rho_{l_{k}}}{\rho_{-k}} \sum_{\substack{\omega,\sigma\\|\omega|=|\sigma|=-l_{k}+1}} \bar{p}_{\sigma}(\bt,2,l_{k}) \bar{p}_{\omega}(\bs,2,l_{k}) \sum_{d=0}^\infty {\beta}^d r_0(-k+1) Term_{(\rho_j,\,j\geq l_{k}+1)}\Big\{\frac{H^d_G(\omega,\sigma)} {r_0(-k+1)}\Big\}\notag\\
  & + \sum_{l_{k+1}=-k}^{0}\frac{\rho_{l_{k+1}}}{\rho_{-k-1}} \sum_{\substack{\omega,\sigma\\|\omega|=|\sigma|=-l_{k+1}}} \bar{p}_{\sigma}(\bt,1,l_{k+1})\bar{p}_{\omega}(\bs,1,l_{k+1}) \sum_{d=0}^\infty {\beta}^d r_0(-k)Term_{(\rho_j,\,j\geq l_{k+1}+1)}\Big\{\frac{H^d_G(\omega,\sigma)} {r_0(-k)}\Big\}\notag\\
= & \cdots\notag\\
= & \frac{\rho_{k}}{\rho_{-1}}\frac{\sum\limits_{\zeta\in S_{k+1}}sgn(\zeta)\mu_{\zeta(1)}^k \cdots\mu_{\zeta(k)}\mu_{\zeta(k+1)}^{-k-1}} {\Delta(\mu)}\frac{\sum\limits_{\tilde{\zeta}\in S_{k+1}}sgn(\tilde{\zeta})\nu_{\tilde{\zeta}(1)}^k \cdots\nu_{\tilde{\zeta}(k)}\nu_{\tilde{\zeta}(k+1)}^{-k-1}}{\Delta(\nu)}\notag\\
  & + \sum_{l_{2}=-1}^{k-1}\frac{\rho_{l_{2}}}{\rho_{-2}} \sum_{\substack{\omega,\sigma\\|\omega|=|\sigma|=k-l_{2}-1}} \bar{p}_{\sigma}(\bt,k-1,l_2) \bar{p}_{\omega}(\bs,k-1,l_2) \sum_{d=0}^\infty {\beta}^d Term_{(\rho_j,\,j\geq l_{2}+1)}\{H^d_G(\omega,\sigma)\}\notag\\
  & + \cdots\notag\\
  & + \sum_{l_{k+1}=-k}^{0}\frac{\rho_{l_{k+1}}}{\rho_{-k-1}} \sum_{\substack{\omega,\sigma\\|\omega|=|\sigma|=-l_{k+1}}} \bar{p}_{\sigma}(\bt,1,l_{k+1})\bar{p}_{\omega}(\bs,1,l_{k+1}) \sum_{d=0}^\infty {\beta}^d r_0(-k) Term_{(\rho_j,\,j\geq l_{k+1}+1)}\Big\{\frac{H^d_G(\omega,\sigma)} {r_0(-k)}\Big\},\notag
\end{align}
where $\bar{p}_{\sigma}(\bt,m,l)=\sum_{\substack{\alpha_1,\cdots,\alpha_m=1\\ \alpha_i\neq\alpha_j,i\neq j}}^{k+1} \frac{\mu_{\alpha_1}^{k}\mu_{\alpha_2}^{k-1} \cdots \mu_{\alpha_{m-1}}^{k+2-m}\mu_{\alpha_m}^{-l-1}} {\prod\limits_{\substack{i=1\\i\neq\alpha_1}}^{k+1}(\mu_{\alpha_1}-\mu_i) \prod\limits_{\substack{i=1\\i\neq\alpha_1,\alpha_2}}^{k+1}(\mu_{\alpha_2}-\mu_i) \cdots \prod\limits_{\substack{i=1\\i\neq\alpha_1,\cdots,\alpha_m}}^{k+1} (\mu_{\alpha_2}-\mu_i)}$\\
$\times p_\sigma(\{\mu_1,\cdots,\hat{\mu}_{\alpha_1},\cdots, \hat{\mu}_{\alpha_m},\cdots,\mu_{k+1}\})$.
\end{proof}
\end{THM}
The $p_\lambda(\bt)$ are a set of bases for the symmetric functions ring $\Lambda(\bt)$\cite{Mac}, and $\bar{p}_{\sigma}(\bt,m,l)$, $\frac{1}{\Delta(\mu)}\sum_{\zeta\in S_{k+1}}$ $sgn(\zeta)\mu_{\zeta(1)}^k \cdots$ $\mu_{\zeta(k)}\mu_{\zeta(k+1)}^{-k-1}$ are symmetric functions belonging to $\Lambda(\bt)$, both of them could be written as a linear combination of $p_\lambda(\bt),\,|\lambda|=k+1$. It is the same for $p_\lambda(\bs)$. If we could get the coefficients of the linear combinations of $p_\lambda(\bt)$ and $p_\lambda(\bs)$, we could write both sides of (\ref{tep34}) as linear combinations of $p_\lambda(\bt)$ and $p_\lambda(\bs)$. Then we could find the recursion formula for the weighted Hurwitz numbers, which means finding the value of $H^d_G(\omega,\sigma),\,|\omega|=|\sigma|=k+1$ by the value of $H^d_G(\omega,\sigma),\,|\omega|=|\sigma|<k+1$. \par

\section{Matrix representation of the Hurwitz numbers}
~\par
For any weighted generating function $G(z)$, the weighted Hurwitz numbers degenerate into the Hurwitz numbers, when $d=0$. That is to say, the Hurwitz numbers correspond to the 0th-order expansion of the weighted generating function.
\begin{equation}\label{tep12-0}
\begin{split} \sum_{\substack{\omega,\sigma\\|\omega|=|\sigma|}}H(\omega,\sigma)p_\omega(\bt) p_\sigma(\mathbf{s}) =\frac{1}{\Delta(\nu)\Delta(\mu)} Term_{\beta^0}\{r_0(-n)\mathop{det} \limits_{i,j=1\cdots n}\Big(\sum_{l=-n}^\infty \nu_j^{-l-1}\mu_i^{-l-1}\rho_l\Big)\},
\end{split}
\end{equation}
where $Term_{\beta^d}\{\cdots\}$ denotes the term of $\beta^d$.\par
Recall the definition of $\rho_j$:
\begin{align}\label{tep11}
 \rho_j = & \prod_{k=1}^j G(\beta k)=\prod_{k=1}^j \sum_{i_k=0}^\infty G_{i_k}(\beta k)^{i_k}, \qquad j>0, \\
 \rho_{-j}= & \prod_{k=1}^{j-1} \frac{1}{G(-\beta k)},\qquad j<0.
\end{align}
The zero-order expansion of $\rho_j,\,j\in\mathbb{Z}$ along the $\beta$ is $1$. Meanwhile
\begin{equation}
\begin{split}
r_0(-n)= & \frac{1}{\rho_{-1}\cdots\rho_{-n}}.
\end{split}
\end{equation}
The zero-order expansion of $r_0(-n)$ along the $\beta$ is also $1$. Assuming $|\nu_j\mu_i|>1$,
\begin{equation}
\begin{split}
  & \sum_{\substack{\omega,\sigma\\|\omega|=|\sigma|}}H^0_G(\omega,\sigma) p_\omega(\bt)p_\sigma(\mathbf{s}){\Delta(\nu)\Delta(\mu)}\\
= & \mathop{det}\limits_{i,j=1\cdots n}\Big(\sum_{l=-n}^\infty \nu_j^{-l-1}\mu_i^{-l-1}\Big)\\
= & \mathop{det}\limits_{i,j=1\cdots n}\Big(\frac{\nu_j^{n}\mu_i^{n}}{\nu_j\mu_i-1}\Big)\\
= & \nu_1^{n}\cdots\nu_n^{n}\mu_1^{n}\cdots\mu_n^{n}\mathop{det}\limits_{i,j=1\cdots n}\Big(\frac{1}{\nu_j\mu_i-1}\Big).\\
\end{split}
\end{equation}
Recalling the Cauchy determinant \cite{Hua} for the sequence , $\{z_i\}_{i=1}^n$$\{w_j\}_{j=1}^n$,
\begin{equation}
det\Big(\frac{1}{z_i-w_j}\Big)=(-1)^{\frac{n(n-1)}{2}}\frac{\Delta(z) \Delta(w)}{\prod_{1\leq i,j\leq n}(z_i-w_j)}.
\end{equation}
Substitute $\mu_i$ into $z_i$ and $\nu_j$ into $\frac{1}{w_j}$, we then have
\begin{equation}
det\Big(\frac{1}{\mu_i\nu_j-1}\Big)=\frac{\Delta(\nu)\Delta(\mu)}{\prod\limits_{1\leq i,j\leq n}(\nu_j\mu_i-1)}
\end{equation}
and
\begin{equation}
\begin{split}
  \sum_{\substack{\omega,\sigma\\|\omega|=|\sigma|}}H^0_G(\omega,\sigma)p_\omega(\bt)p_\sigma(\mathbf{s})
= & \frac{\nu_1^{n}\cdots\nu_n^{n}\mu_1^{n}\cdots\mu_n^{n}}{\prod\limits_{1\leq i,j\leq n}(\nu_j\mu_i-1)}.\\
\end{split}
\end{equation}\par
Notice that $|\nu_j\mu_i|>1$, we have $\frac{1}{\nu_j\mu_i-1}=\sum\limits_{k=-\infty}^{-1}\nu_j^{k}\mu_i^{k}$ and
\begin{equation}\label{tep13}
\begin{split}
  \sum_{\substack{\omega,\sigma\\|\omega|=|\sigma|}}H^0_G(\omega,\sigma)p_\omega(\bt)p_\sigma(\mathbf{s})
= &~ \nu_1^{n}\cdots\nu_n^{n}\mu_1^{n}\cdots\mu_n^{n}\prod\limits_{1\leq i,j\leq n}\sum\limits_{k_{ij}=-\infty}^{-1}\nu_j^{k}\mu_i^{k}\\
= & \prod\limits_{1\leq i,j\leq n}\sum\limits_{k_{ij}=-\infty}^{0}\nu_j^{k}\mu_i^{k}\\
= & \sum_{k_{11}\cdots k_{nn}=-\infty}^{0}\prod\limits_{1\leq j\leq n}\nu_j^{\sum_{i=1}^{n}k_{ij}}\prod\limits_{1\leq i \leq n}\mu_i^{\sum_{j=1}^{n}k_{ij}}.
\end{split}
\end{equation}\par
$\prod\limits_{1\leq j\leq n}\nu_j^{\sum_{i=1}^{n}k_{ij}}\prod\limits_{ 1\leq i \leq n}\mu_i^{\sum_{j=1}^{n}k_{ij}}$ directly corresponds to an $n\times n$-dimensional square matrix,
\begin{equation}\label{tep14}
\left(\begin{array}{cccc}
k_{11} & k_{12} & \cdots & k_{1n}\\
k_{21} & k_{22} & \cdots & k_{2n}\\
\vdots & \vdots & \ddots & \vdots\\
k_{n1} & k_{n2} & \cdots & k_{nn}
\end{array}\right),
\end{equation}
the added value of $\{k_{ij}\}_{i=1}^n$ in each column is the power of $\nu_j$ (the added value of $\{k_{ij}\}_{j= 1}^n$ in each row is the power of $\mu_i$.) $\nu_1^{m_1}\cdots\nu_{n}^{m_n}\mu_1^{M_1}\cdots\mu_n^{M_n}$ may correspond to more than one matrix, like (\ref{tep14}). In this way, it establishes a semigroup homomorphism from an $n \times n$-dimensional non-positive integer coefficient matrix semigroup with no inverse element to monomial semigroups of $\nu_1,\,\cdots,\,\nu_n,\,\mu_1,\,\cdots, \,\mu_n$.
\begin{equation}\nonumber
\begin{split}
\{Mn(n,\mathbb{Z}_-),\,+\} & \longrightarrow \{\{\nu_1^{m_1}\cdots\nu_n^{m_n}\mu_1^{M_n}\mu_{n}^{M_n}|m_1,\cdots,m_n,M_1\cdots,M_n\in \mathbb{Z}_-\},\,\times\}, \\
\left(\begin{array}{cccc}
k_{11} & k_{12} & \cdots & k_{1n}\\
k_{21} & k_{22} & \cdots & k_{2n}\\
\vdots & \vdots & \ddots & \vdots\\
k_{n1} & k_{n2} & \cdots & k_{nn}
\end{array}\right)                  & \longrightarrow  \nu_1^{m_1}\cdots\nu_{n}^{m_n}\mu_1^{M_1}\cdots\mu_n^{M_n},
\end{split}
\end{equation}
in which $\mathbb{Z}_-=\{z\in\mathbb{Z}|\,z\leq 0\}$ is a non-positive integer. The group addition of the $n\times n$ dimensional non-positive integer coefficient matrix semigroup  is matrix addition, and the group multiplication of the monomial semigroups of $\nu_1,\,\cdots,\,\nu_n,\,\mu_1,\,\cdots,\,\mu_n$ is multiplication. \par
Similarly, $p_\omega(\bt)p_\sigma(\mathbf{s})$ could be decomposed into the sum of $n\times n$-dimensional non-positive integer coefficient matrix groups,
\begin{equation}
\bordermatrix{
       & m_1    & m_2    & \cdots & m_n\cr
M_1    & k_{11} & k_{12} & \cdots & k_{1n}\cr
M_2    & k_{21} & k_{22} & \cdots & k_{2n}\cr
\vdots & \vdots & \vdots & \ddots & \vdots\cr
M_n    & k_{n1} & k_{n2} & \cdots & k_{nn}},
\end{equation}
with $\{m_1,m_2,\cdots,m_n\}=\{\omega_1,\omega_2,\cdots,\omega_n\}$ and $\{M_1,M_2,\cdots,M_n\}=\{\sigma_1,\sigma_2,\cdots,\sigma_n\}$. In this way, we obtain the matrix representation of the Hurwitz numbers. We could calculate the value of the Hurwitz numbers by matrix operations.\par
When $d\neq 0$, the value of $H^d_G(\omega,\sigma)$ depends on the construction of the weighted generating function $G(z)$.

\section{Determinant representation of the weighted paths in Cayley graph}
~\par
M. Guay-Paquet and J. Harnad investigated a parametric family of 2D Toda $\tau$ functions \cite{PJGen,PJ2D}, which are the generating functions of the weighted Hurwitz numbers, and the weighted paths are defined by the Cayley graph of the symmetric group $\mathbf{S}_n$. We obtained the determinant representation of the 2D Toda $\tau$ functions in section 3, and then in the next section, we obtained the determinant representation of the weighted Hurwitz numbers. In this section, we will carry out the determinant representation of the weighted paths in the Cayley graph of the symmetric group $\mathbf{S}_n$.\par
Recall that $m^\lambda_{\mu\nu}$ is the number of monotonic $\lambda$ signature paths in the Cayley graph from $cyc(\mu)$ to $cyc(\nu)$ with in (\ref{Cayley})
\begin{equation}\nonumber
H^d_G(\mu,\nu)=F^d_G(\mu,\nu)=\frac{1}{|\nu|!}\sum\limits_{\lambda,|\lambda|=d} G_\lambda m^\lambda_{\mu\nu}.
\end{equation}
To get the determinant representation of the weighted paths in the Cayley graph, we need to expand the weighted Hurwitz numbers along $\beta$.
\begin{equation}\label{tep12-1}
\begin{split} \sum_{\substack{\omega,\sigma\\|\omega|=|\sigma|}}H^d_G(\omega,\sigma)p_\omega(\bt) p_\sigma(\mathbf{s}) =\frac{1}{\Delta(\nu)\Delta(\mu)} Term_{\beta^d}\Big\{r_0(-n)\mathop{det} \limits_{i,j=1\cdots n}\Big(\sum_{l=-n}^\infty \nu_j^{-l-1}\mu_i^{-l-1}\rho_l\Big)\Big\},
\end{split}
\end{equation}
where $Term_{\beta^d}\{\cdots\}$ denotes the term of $\beta^d$. From (\ref{Cayley}) and (\ref{tep12-1}),
\begin{equation}
\begin{split}
\sum_{\substack{\omega,\sigma\\|\omega|=|\sigma|=n}}  & \frac{1}{n!} \sum\limits_{\lambda,|\lambda|=d} G_\lambda m^\lambda_{\mu\nu}p_\omega(\bt) p_\sigma(\mathbf{s})\\
= & Term_{-n}\Big\{\frac{1}{\Delta(\nu)\Delta(\mu)} Term_{\beta^d}\Big\{r_0(-n)\mathop{det} \limits_{i,j=1\cdots n}\Big(\sum_{l=-n}^\infty \nu_j^{-l-1}\mu_i^{-l-1}\rho_l\Big)\Big\}\Big\}.
\end{split}
\end{equation}
By Lemma [\ref{LEM:N_P}], we have
\begin{align}\label{tep36}
\sum_{\substack{\omega,\sigma\\|\omega|=|\sigma|=n}}  & \frac{1}{n!} \sum\limits_{\lambda,|\lambda|=d} G_\lambda m^\lambda_{\mu\nu}p_\omega(\bt) p_\sigma(\mathbf{s})\\
= & \frac{1}{\Delta(\nu)\Delta(\mu)} \sum_{\substack{l_1,\cdots,l_{n}=-n\\l_1>\cdots>l_n\\l_1+\cdots +l_n=\frac{-n(n-1)}{2}}}^{\infty} Term_{\beta^d}\Big\{\frac{\rho_{l_1}\cdots\rho_{l_n}} {\rho_{-1}\cdots\rho_{-n}}\Big\}\notag\\
  & \sum_{\sigma\in S_k} sgn(\sigma)\left|\begin{array}{cccc}
\nu_{\sigma(1)}^{-l_1-1}\mu_{1}^{-l_1-1} & \nu_{\sigma(2)}^{-l_2-1}\mu_{1}^{-l_2-1} & \cdots & \nu_{\sigma(N)}^{-l_n-1}\mu_{1}^{-l_n-1}\\
\nu_{\sigma(1)}^{-l_1-1}\mu_{2}^{-l_1-1} & \nu_{\sigma(2)}^{-l_2-1}\mu_{2}^{-l_2-1} & \cdots & \nu_{\sigma(N)}^{-l_n-1}\mu_{2}^{-l_n-1}\\
\vdots & \vdots & \ddots & \vdots\\
\nu_{\sigma(1)}^{-l_1-1}\mu_{N}^{-l_1-1} & \nu_{\sigma(2)}^{-l_2-1}\mu_{N}^{-l_2-1} & \cdots & \nu_{\sigma(N)}^{-l_n-1}\mu_{N}^{-l_n-1}\\
\end{array}\right|.\notag
\end{align}
Only $\rho_j,\,j\in\mathbb{Z}$ on the right side of (\ref{tep36}) are functions of $G_\lambda$. Let us focus on $\frac{\rho_{l_1}\cdots\rho_{l_n}} {\rho_{-1}\cdots\rho_{-n}},\,l_1>\cdots>l_n\geq-n,\,l_1+\cdots +l_n=\frac{-n(n-1)}{2}$. We have $n-1\geq l_1\geq-1,\,l_{1}>l_2\geq-2\cdots,\,l_{n-1}>l_n\geq-n$ and
\begin{align}
\frac{\rho_{l_i}}{\rho_{-i}}= & r_{l_i}r_{l_i-1}\cdots r_{-i+1}\\
                            = & G(\beta l_i)G(\beta l_{l_i-1})\cdots G( \beta (-i+1))\notag\\
                            = & \sum_{k_{i,l_i}=0}^\infty G_{k_{i,l_i}}(\beta l_i)^{k_{i,l_i}}\sum_{k_{i,l_i-1}=0}^\infty G_{k_{i,l_i-1}}(\beta l_i-1)^{k_{i,l_i-1}}\notag\\
                              & \cdots \sum_{k_{i,{-i+1}}=0}^\infty G_{k_{i,{-i+1}}}(\beta (-i+1))^{k_{i,{-i+1}}}.\notag
\end{align}
If $l_i=-i$, $\frac{\rho_{l_i}}{\rho_{-i}}=1$. Notice that $l_1+\cdots +l_n=\frac{-n(n-1)}{2}$, then $l_1-(-1)+l_2-(-2)+\cdots+l_n-(-n)=n$, that is to say there are $n$
numbers in the set $\{l_1,l_1-1, \cdots,0,l_2,l_2-1,\cdots,-1,\cdots,l_n,l_n-1,\cdots,-n+1\}$. We have
\begin{align}\label{tep37}
  & Term_{\beta^d}\Big\{\frac{\rho_{l_1}\cdots\rho_{l_n}} {\rho_{-1}\cdots\rho_{-n}}\Big\}\\
= & Term_{\beta^d}\Big\{ \sum_{k_{1,l_1}=0}^\infty G_{k_{1,l_1}}(\beta l_1)^{k_{1,l_1}}\sum_{k_{1,l_1-1}=0}^\infty G_{k_{1,l_1-1}}(\beta l_1-1)^{k_{1,l_1-1}}\cdots\notag\\
  & \sum_{k_{1,{0}}=0}^\infty G_{k_{1,{0}}}(\beta (0))^{k_{i,0}} \cdots \sum_{k_{n,l_n}=0}^\infty G_{k_{n,l_n}}(\beta l_n)^{k_{n,l_n}}\sum_{k_{n,l_n-1}=0}^\infty G_{k_{n,l_n-1}}\notag\\
  & \times(\beta l_n-1)^{k_{n,l_n-1}}\cdots \sum_{k_{n,{-n+1}}=0}^\infty G_{k_{n,{-n+1}}}(\beta (-n+1))^{k_{n,{-n+1}}}\Big\}\notag\\
= & \sum_{\substack{\lambda,|\lambda|=d,\\l(\lambda)\leq n}}G_{\lambda}(l_1,l_1-1, \cdots,0,l_2,l_2-1,\cdots,-1,\notag\\
  & \cdots,l_n,l_n-1,\cdots,-n+1)^{\lambda}\beta^d.\notag
\end{align}
where $\mu^\lambda=\sum_{\sigma\in \mathbf{S}_n}\mu_1^{\lambda_{\sigma(1)}}\cdots\mu_n^{\lambda_{\sigma(n)}}$ is an abbreviation mentioned in (\ref{eq:abbreviate})\cite{Mac}. Substituting (\ref{tep37}) into (\ref{tep36}), we have
\begin{align}\label{tep38}
\sum_{\substack{\omega,\sigma\\|\omega|=|\sigma|=n}}  & \frac{1}{n!} \sum\limits_{\lambda,|\lambda|=d} G_\lambda m^\lambda_{\mu\nu}p_\omega(\bt) p_\sigma(\mathbf{s})\\
= & \frac{1}{\Delta(\nu)\Delta(\mu)} \sum_{\substack{l_1,\cdots,l_{n}=-n\\l_1>\cdots>l_n\\l_1+\cdots +l_n=\frac{-n(n-1)}{2}}}^{\infty} \sum_{\substack{\lambda,|\lambda|=d,\\ l(\lambda)\leq n}} G_{\lambda}(l_1,l_1-1, \cdots,0,\notag\\
  & l_2,l_2-1,\cdots,-1,\cdots,l_n,l_n-1,\cdots,-n+1)^{\lambda}\beta^d\notag\\
  & \sum_{\sigma\in S_k} sgn(\sigma)\left|\begin{array}{cccc}
\nu_{\sigma(1)}^{-l_1-1}\mu_{1}^{-l_1-1} & \nu_{\sigma(2)}^{-l_2-1}\mu_{1}^{-l_2-1} & \cdots & \nu_{\sigma(N)}^{-l_n-1}\mu_{1}^{-l_n-1}\\
\nu_{\sigma(1)}^{-l_1-1}\mu_{2}^{-l_1-1} & \nu_{\sigma(2)}^{-l_2-1}\mu_{2}^{-l_2-1} & \cdots & \nu_{\sigma(N)}^{-l_n-1}\mu_{2}^{-l_n-1}\\
\vdots & \vdots & \ddots & \vdots\\
\nu_{\sigma(1)}^{-l_1-1}\mu_{N}^{-l_1-1} & \nu_{\sigma(2)}^{-l_2-1}\mu_{N}^{-l_2-1} & \cdots & \nu_{\sigma(N)}^{-l_n-1}\mu_{N}^{-l_n-1}\\
\end{array}\right|.\notag
\end{align}
Since $m^\lambda_{\mu\nu}$ is the number of monotonic $\lambda$ signature paths in the Cayley graph of $\mathbf{S}_n,\,|\mu|=|\nu|=n$ from $cyc(\mu)$ to $cyc(\nu)$, $m^\lambda_{\mu\nu}=0,\,l(\lambda)>n$. The left side of (\ref{tep38}) does not emphasize $l(\lambda)\leq n$. We find the determinant representation of the weighted paths in
the Cayley graph,
\begin{align}\label{tep39}
\sum_{\substack{\omega,\sigma\\|\omega|=|\sigma|=n}}  & m^\lambda_{\mu\nu}p_\omega(\bt) p_\sigma(\mathbf{s})\\
= & \frac{n!}{\Delta(\nu)\Delta(\mu)} \sum_{\substack{l_1,\cdots,l_{n}=-n\\l_1>\cdots>l_n\\l_1+\cdots +l_n=\frac{-n(n-1)}{2}}}^{\infty} (l_1,l_1-1, \cdots,0,l_2,l_2-1,\cdots,\notag\\
  & -1,\cdots,l_n,l_n-1,\cdots,-n+1)^{\lambda}\beta^d\notag\\
  & \sum_{\sigma\in S_k} sgn(\sigma)\left|\begin{array}{cccc}
\nu_{\sigma(1)}^{-l_1-1}\mu_{1}^{-l_1-1} & \nu_{\sigma(2)}^{-l_2-1}\mu_{1}^{-l_2-1} & \cdots & \nu_{\sigma(N)}^{-l_n-1}\mu_{1}^{-l_n-1}\\
\nu_{\sigma(1)}^{-l_1-1}\mu_{2}^{-l_1-1} & \nu_{\sigma(2)}^{-l_2-1}\mu_{2}^{-l_2-1} & \cdots & \nu_{\sigma(N)}^{-l_n-1}\mu_{2}^{-l_n-1}\\
\vdots & \vdots & \ddots & \vdots\\
\nu_{\sigma(1)}^{-l_1-1}\mu_{N}^{-l_1-1} & \nu_{\sigma(2)}^{-l_2-1}\mu_{N}^{-l_2-1} & \cdots & \nu_{\sigma(N)}^{-l_n-1}\mu_{N}^{-l_n-1}\\
\end{array}\right|.\notag
\end{align}

\bigskip
%%%%%%%%%%%%%%%%%%%%% Acknowledgements??D? %%%%%%%%%%%%%%%%%
\bigskip
\noindent
{\small \it Acknowledgements.}The financial supports from the Natural Science Foundation of China (NSFC, Grants 11775299) and National Key Research and Developing Program of China (NKRDPC, Grants 2018YFB0704304) are gratefully acknowledged from one of the authors (Ding).

\bigskip

%%%%%%%%%%%%%%%% Bibliography2???????¨¢ %%%%%%%%%%%%%%%%

%\newcommand{\arxiv}[1]{\href{http://arxiv.org/abs/#1}{arXiv:{#1}}}

%\begin{thebibliography}{99}

\end{document}